\documentclass[iop,numberedappendix]{emulateapj}

\usepackage{natbib,amsmath}

\usepackage{graphicx}
\usepackage{natbib}
\usepackage{amssymb}
\usepackage{float}
\usepackage{makeidx}
\usepackage{amsmath}
\usepackage{rotating}
\usepackage{longtable}
\usepackage{subfigure}
\usepackage{morefloats}
\usepackage{captcont}
\usepackage{xcolor} 
\usepackage{chngcntr}

\nocite{*}

\graphicspath { {./}}

\def\ts{\thinspace}

\newcommand{\etal}{et~al.\/}

\newcommand{\cm}{cm$^{-2}$ }

\newcommand\mlstar{L^*}
\newcommand\lstar{$\mlstar$}

\shortauthors{Bish \etal}
\shorttitle{Galactic Gas Flows from Halo to Disk}

\begin{document}

\title{Galactic Gas Flows from Halo to Disk:\\Tomography and Kinematics at the Milky Way's Disk-Halo Interface}

\author{Hannah V. Bish\altaffilmark{1}, 
Jessica K. Werk\altaffilmark{1,2}, 
J. Xavier Prochaska\altaffilmark{2}, 
Kate H.R. Rubin\altaffilmark{3}, 
Yong Zheng\altaffilmark{4,5}, \\
John M. O'Meara\altaffilmark{6}, and
Alis J. Deason\altaffilmark{7} 
}

\altaffiltext{1}{Department of Astronomy, University of Washington, Seattle, WA 98195, USA; $hvbish@uw.edu$}
\altaffiltext{2}{UCO/Lick Observatory, University of California, Santa Cruz, CA 95064, USA}
\altaffiltext{3}{Department of Astronomy, San Diego State University, San Diego, CA 92182, USA}
\altaffiltext{4}{Department of Astronomy, University of California, Berkeley, USA}
\altaffiltext{5}{Miller Institute for Basic Research in Science, University of California, Berkeley, USA}
\altaffiltext{6}{W. M. Keck Observatory, 65--1120 Mamalahoa Highway, Kamuela, HI 96743, USA}
\altaffiltext{7}{Institute for Computational Cosmology, Department of Physics, University of Durham, South Road, Durham DH1 3LE, UK}

\begin{abstract}

We present a novel absorption line survey using 54 blue horizontal branch stars (BHBs) in the Milky Way halo as background sources for detecting gas flows at the disk-halo interface. Distance measurements to high-latitude ($b$ $>$ 60$^{\circ}$) background stars at 3.1 - 13.4 kpc, combined with unprecedented spatial sampling and spectral resolution, allow us to examine the 3-dimensional spatial distribution and kinematics of gas flows near the disk. We detect absorption signatures of extraplanar CaII and NaI in Keck HIRES spectra and find that their column densities exhibit no trend with distance to the background sources, indicating that these clouds lie within 3.1 kpc of the disk. We calculate covering fractions of $f_{\rm CaII} = 63\%$, $f_{\rm NaI} = 26\%$, and $f_{\rm HI} = 52\%$, consistent with a picture of the CGM that includes multi-phase clouds containing small clumps of cool gas within hotter, more diffuse gas. Our measurements constrain the scale of any substructure within these cool clouds to $< 0.5$ kpc. CaII and NaI absorption features exhibit an intermediate-velocity (IV) component inflowing at velocities of $-75$ km/s $< v < -25$ km/s relative to the local standard of rest, consistent with previously-studied HI structures in this region. We report the new detection of an inflow velocity gradient $\Delta v_z \sim 6-9$ km/s/kpc across the Galactic plane. These findings place constraints on the physical and kinematic properties of CGM gas flows through the disk-halo interface, and support a galactic fountain model in which cold gas rains back onto the disk.\\

\end{abstract}


\section{Introduction}
\label{sec:intro}

Galaxies like the Milky Way grow via complex processes balancing the supply, consumption, and removal of gas in star-forming regions. Observations of the baryons involved in this cycle are consistent with gas moving in a ``galactic fountain" \citep{Bregman1980}, in which hot gas is ejected from the disk via supernovae, stellar winds, and/or AGN, then cools and rains back down onto the disk to trigger new star formation \citep{shapirofield1976,savagesembach1994,sembach2003}. This continual cycling of baryons between the halo and disk plays a crucial role in regulating the supply of fuel for star formation and consequently has a major influence on galaxy evolution.

The varied and complex dynamic mechanisms which drive Galactic baryon cycling are not fully understood, but may be key to answering long-standing questions about how galaxies sustain star formation, why they become quenched, and the baryons and metals that are `missing' from the disks of star-forming galaxies \citep{tumlinsonARAA2017}. The relative contributions of  star-formation driven winds in the disk, a biconical wind from Galactic center \citep{su2010}, tidal stripping of satellites, and inflowing gas from the intergalactic medium \citep{brooks2009} are still widely debated \citep{bordoloi2014b,fox2015a}.

The diffuse halo extending from the roughly-defined boundary of a galaxy's disk out to its virial radius -- the circumgalactic medium (CGM) -- is host to the galactic fountain and serves as a massive reservoir of gas ($>10^{10}$ $\rm M_{\odot}$ for \lstar{} galaxies; \citealt{werk2014,prochaska2017,stern2016,stocke2013}). However, observations of the CGM are challenging because the gas it contains is approximately a million times less dense than the ISM and therefore difficult to detect in emission for single $L^*$ Milky Way-like galaxies at all epochs. \citep{stocke2013,werk2014}. Early evidence for its existence emerged when \cite{munch1961} identified NaI and CaII absorption in the spectra of high-latitude Milky Way stars as a signature of extraplanar gas. 

Subsequent work identified halo gas absorption in external galaxies using quasars as background sources \citep{bahcallspitzer1969,bergeron1986}. More recently, deep HI-21cm observations of gas at the disk-halo interface of both the Milky Way and other external, edge-on spiral galaxies have found a so-called halo lag \citep[e.g.]{putman2012}. The approximate rate of halo gas rotation decreases with z-height, with a typical velocity drop-off of $15-30$ km/s/kpc \citep[e.g.]{fraternali2002,sancisi2008, heald2011}. This observed halo lag has been offered as evidence of an infalling gaseous component of the disk-halo interface \citep{fraternalibinney2006}.

Gas at the disk-halo interface is of particular interest because of the rich physical information it contains about material which has cycled through the CGM.
Constraints on the content, kinematics, and spatial extent of this material can ultimately point to its origin, but direct detection via emission is possible only for the densest material in the Milky Way halo. Full-sky HI 21-cm emission maps reveal a population of neutral intermediate-velocity clouds (IVCs, 25 $<$ $|v|$ $<$ 90 km/s relative to the local standard of rest) at heights of $\sim$0.5-3 kpc above the disk \citep{kuntzdanly1996,wakker2001,smoker2011}. IVCs are found at predominantly negative velocities \citep{wakker2001}, and like high-velocity clouds in the halo, their gas is multi-phase \citep{albertdanly2004,richter2017,lehnerhowk2011,fox2014}. They are thought to contain material that once originated in the Galaxy's disk because their metallicities are consistent with that of Milky Way ISM gas, and because ISM simulations find that IVC morphology and kinematics are reproduced by ISM ejecta \citep{Bregman1980, deAvillez2000, ford2010, richter2017}.

The timescales on which this material accretes, and the phase structure of the clouds themselves, are two of the physical properties that can best constrain models of the galactic fountain. However, isolating halo gas from disk gas at the interface is challenging because line-of-sight observations themselves do not provide distance information about the gas, and the two components may overlap partially in velocity space.

Within the last decade, significant advances have been made with improved spectral analysis techniques and the installation of the Cosmic Origins Spectrograph (COS) on the Hubble Space Telescope, which increased the sensitivity of diffuse gas detections by an order of magnitude \citep{osterman2009}. Today, the technique most sensitive to low-density gas along individual sightlines uses spectra of background sources to identify absorption features from foreground gas. These background sources may be quasars observed through the CGM of foreground galaxies (e.g. \citealt{werk2013,zheng2017b}) or halo stars behind Milky Way CGM gas (e.g. \citealt{savage2009,lehnerhowk2011,fox2015a}). Additionally, the ``down-the-barrel" technique measures a galaxy's CGM in absorption using its own starlight as a background source, and has been used to study outflows and accretion (e.g. \citealt{rubin2012,heckman2015}).

We present an absorption-line survey designed to examine the kinematics, spatial extent, and cloud characteristics of gas at the Milky Way's disk-halo interface. We have analyzed CaII ($\lambda\lambda$ 3934, 3969) and NaI ($\lambda\lambda$ 5891, 5897) transitions in absorption along 54 sightlines to blue horizontal branch (BHB) stars with well-constrained heights 3.1 - 13.4 kpc above the disk. Because they are approximate standard candles, BHBs provide distances which we use to describe the 3-D spatial extent of the gas. The sightlines are limited to high Galactic latitudes ($b$ $>$ 60$^{\circ}$) so that radial velocity measurements place the greatest possible constraints on gas inflow/outflow velocities. The excellent resolution of Keck HIRES spectra and the unprecedented 3-dimensional spatial sampling of this survey allow us to place empirical constraints on the physical and kinematic properties of CGM gas flows through the disk-halo interface.

This paper is organized as follows: in \S\ref{sec:observations}, we describe the survey design and observations. In \S\ref{sec:analysis} we describe our methods for line profile fitting and define subsamples for analysis. In \S\ref{sec:results} we present the column density and velocity measurements of the gas absorbers in this sample. In \S\ref{sec:discussion} we examine how these measurements can place constraints on cloud size and distribution and the implications of our findings for potential origins of the gas we see. In \S\ref{sec:summary} we summarize our findings.
\\

\begin{figure}[b]
    \centering
    \vspace{15pt}
    \includegraphics[width=0.5\textwidth]{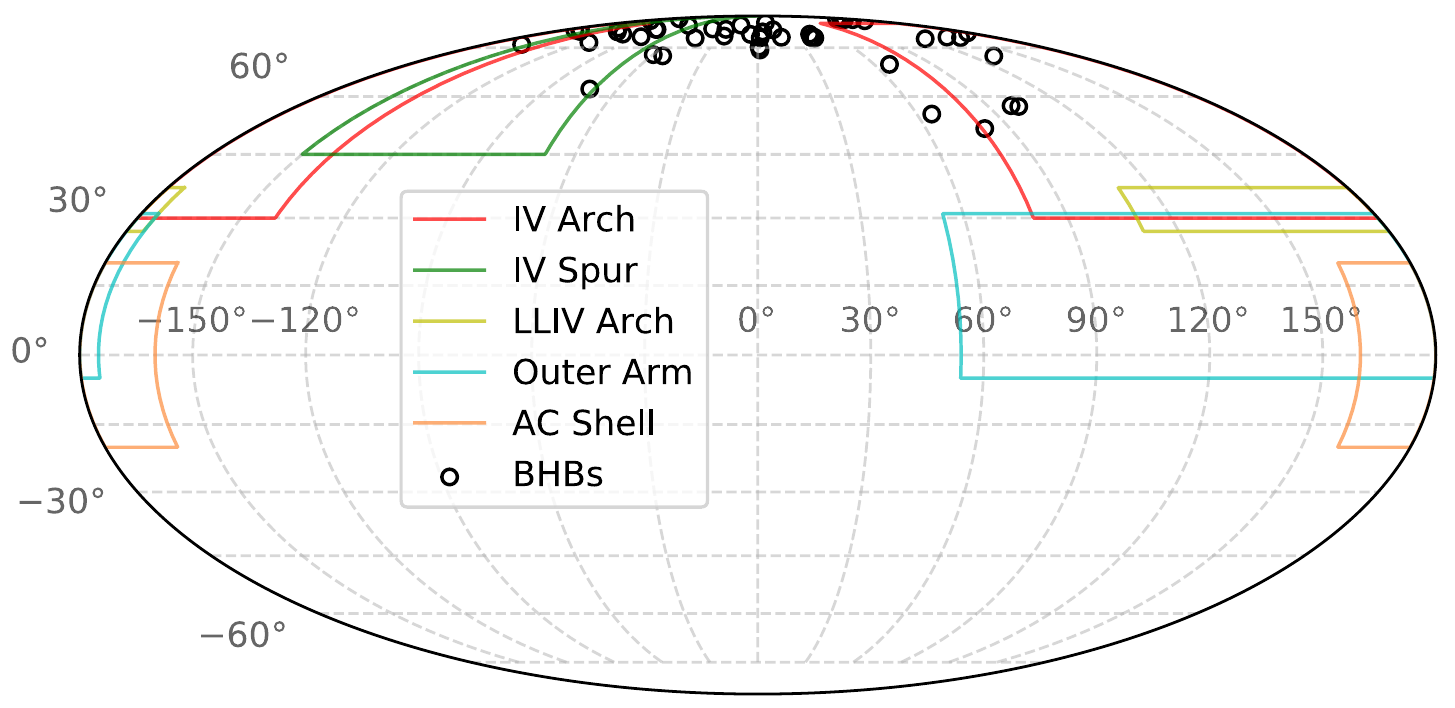}
    \caption{{\sc BHB sightlines \& nearby IVCs.} Galactic coordinates of the observed BHB sightlines on an all-sky map, along with the approximate boundaries of several large intermediate-velocity HI complexes \citep{vanwoerden2005}. The plane of the Milky Way extends horizontally, with Galactic center at $l$$=$0$^\circ$, $b$$=$0$^\circ$. The sightlines probe the boundaries of the high-latitude IVCs known as the IV Arch and the IV Spur.  \\} 
    \label{fig:bhbivc}
\end{figure}

\section{Observations}
\label{sec:observations}

\subsection{Survey Design \& Sample Selection}
Our absorption line survey uses blue horizontal branch stars in the Milky Way halo (hereafter referred to as BHBs) as background sources to study diffuse foreground gas. BHBs are well-suited for this purpose because they have bright, relatively featureless spectra ideal for absorption-line analysis, and as approximate standard candles they have measured distances to within $\pm 10\%$ \citep{kinman1994,sirko2004,clewley2006,xue2008,deason2011}. We have taken advantage of these properties to obtain kinematic information and distance constraints for gas along sightlines which probe the disk-halo interface of the Milky Way. We observed 56 out of 83 stars meeting the selection criteria described in this section, and ultimately excluded two from our analysis. Figure \ref{fig:bhbivc} shows the Galactic coordinates of the sightlines on an all-sky map, along with the approximate boundaries of several large intermediate-velocity HI complexes \citep{vanwoerden2005}. A full list of these targets and their properties can be found in Table \ref{table:bhblist}.

\begin{figure}[t]
	\vspace{15pt}
    \centering
    \hspace{-15pt}
    \includegraphics[width=0.5\textwidth]{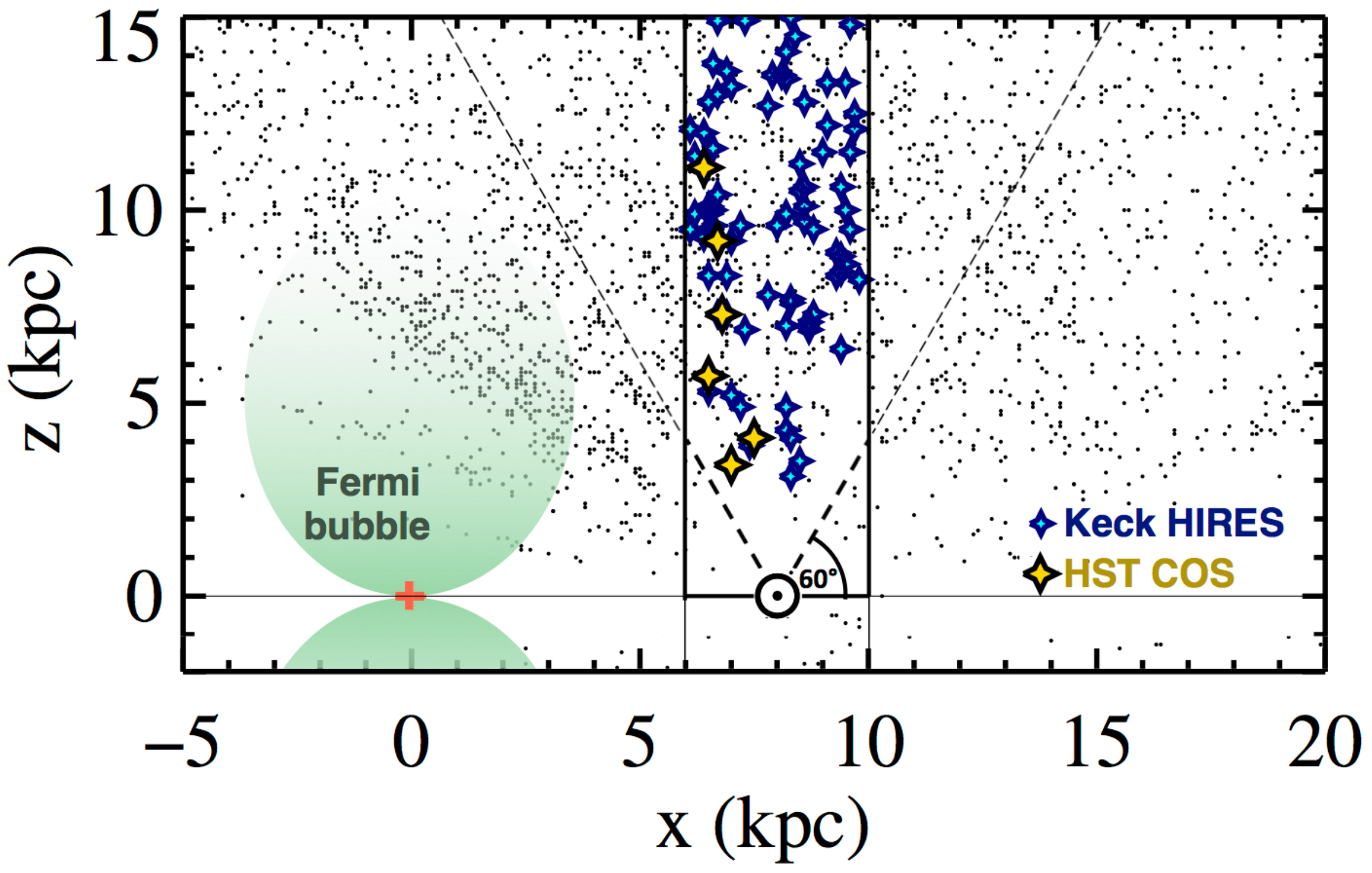} 
    \caption{{\sc Survey geometry.} Spatial distribution of BHB stars in the northern Milky Way halo, shown in the $x$-$z$ projection. The full BHB catalog from SEGUE (\citealt{xue2011}) is displayed with black points. Targets observed with Keck HIRES for this study are highlighted with star symbols, and were selected to fall at high Galactic latitudes within a cylinder 2 kpc in radius centered on the Sun. Yellow stars indicate BHBs which have also been observed with HST COS as part of a complementary program to study highly-ionized gas in this regime. The red cross marks the Galactic center, and the Sun is marked by the solar symbol. Green shading indicates the approximate extent of Fermi bubbles formed by winds emanating from the Galactic center (\citealt{su2010}). The three targets at $b>60^{\circ}$ are included here, but do not appear below the dashed line at $b=60^{\circ}$ as viewed from this particular angle.\\
    }
    \label{fig:keckoverview}
\end{figure}

The sample of background sources was selected from $>$4,500 BHB stars with accurate distances and velocities in the SDSS SEGUE catalog \citep{xue2011}. To verify that we are using the best available distance measurements for the BHBs, we checked the SEGUE distances against the Gaia DR2 catalogue. We found that for all but one BHB, SEGUE distances were more precise because the stars are close to or beyond the $\sim$5 kpc limiting distance for accurate parallax measurement \citep{gaiadr2}. Since the distance measurements we use only set an upper limit on the distance to the absorbing gas, and most of the gas likely sits close to the disk at small distances (see \S \ref{sec:columndensity}), the error on BHB distance measurements does not significantly affect our analysis.

Figure \ref{fig:keckoverview} shows the distribution of SEGUE BHBs from an edge-on view of the Milky Way. The full set of BHBs in the catalog is marked by small black points, and blue stars show the location of BHBs selected for this study. A subset of these sightlines, shown in yellow, have also been observed with HST COS in the UV and data will be presented in Werk et al. 2019 (in prep). The Galactic center is indicated with a red cross, and the approximate extent of the Fermi Bubble is shown in green \citep{su2010}. The Fermi bubble has been probed in absorption by \cite{bordoloi2017} using halo stars with known distances, and other work has examined the Fermi Bubble as a potential source of recycled gas observed elsewhere in the Milky Way \citep{fox2015a,karim2018}.

\begin{figure}[t]
	\vspace{-5pt}
    \centering
    \vspace{20pt}
    \includegraphics[width=0.48\textwidth]{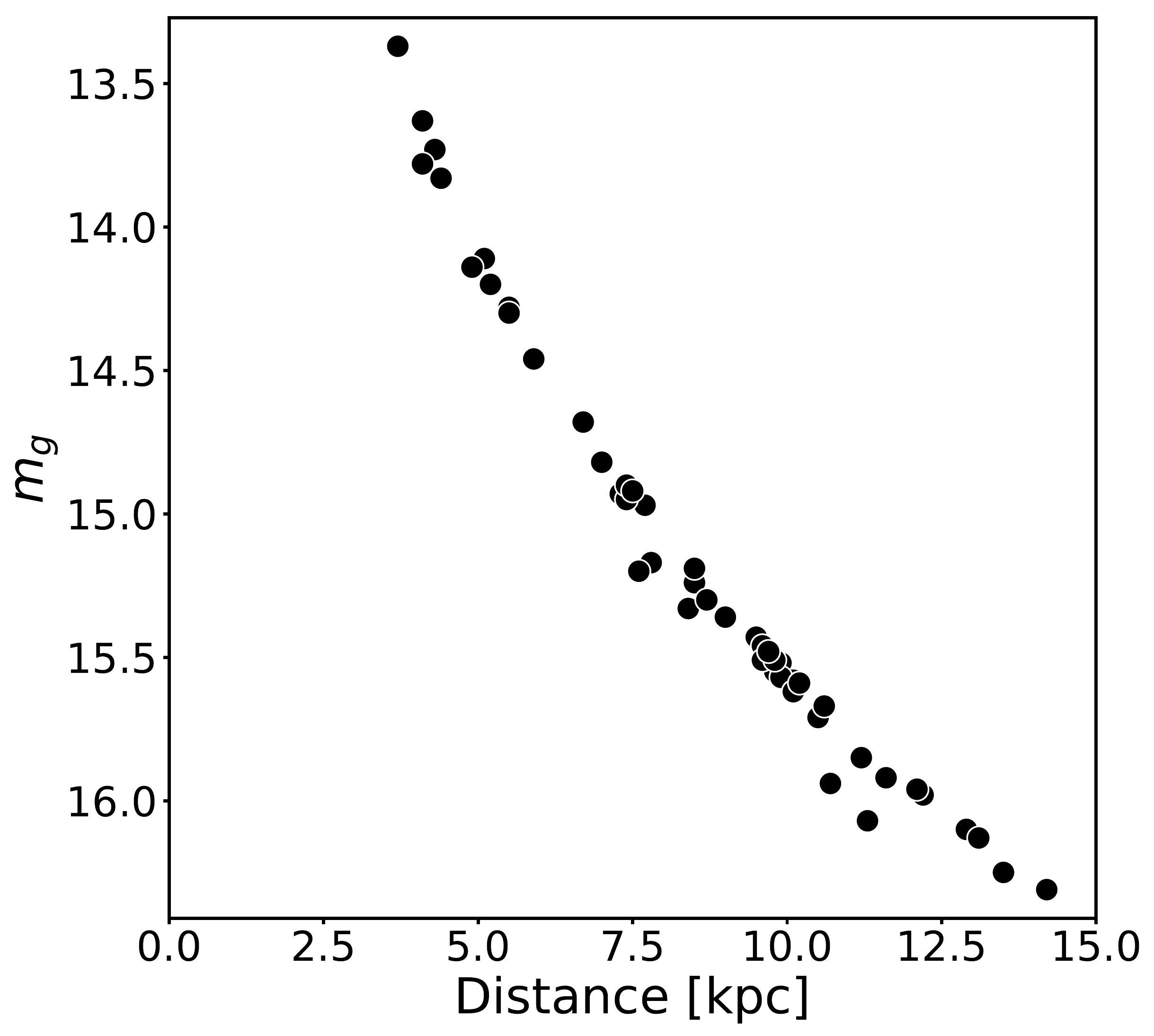}
    \caption{{\sc Background source magnitudes.}  The SDSS apparent $g$-band magnitude of each background source and its distance from the Sun. The sources are blue horizontal branch (BHB) stars, which are approximate standard candles and can be used to constrain the distance to foreground gas along each sightline. Distance calibrations are dependent on both $g$ magnitude and $(g-r)$ color \citep{xue2008}. \\
    }
    \label{fig:gmags}
\end{figure}

This sample meets several selection criteria. First, targets are bright enough to reach S/N $\geq 10$ with Keck/HIRES in less than $\sim 1$\,hr so that absorption features of interest can be detected. This constraint imposes a $g$-band magnitude cut of $m_g < 16.5$. Figure \ref{fig:gmags} shows the distribution of $g$-band magnitudes for background sources in this work and their distances from the Sun. The distance calibrations are dependent on both $g$ magnitude and $(g-r)$ color \citep{xue2008}. Recently, \cite{lancaster2018} found that roughly 10\% of stars in the SDSS BHB sample were contaminant Blue Straggler stars, which are fainter than BHBs and would lead to overestimated distances. We have applied their color cut to the BHB catalog to remove these contaminants.

Second, we excluded background BHBs with radial velocities $-80$ km/s $< v_{\rm star} <$ 0 km/s so that absorption features of the background star and foreground gas would not overlap. This constraint was determined after an initial pilot study showed that virtually all foreground gas fell within that velocity range (see \S\ref{sec:kinematics} for discussion of these results).

To place robust constraints on the kinematics of gas inflows/outflows, we selected high-latitude sightlines so that measured radial velocities will best constrain the motion of this gas. High-latitude sightlines provide a window into the dynamics of extraplanar gas flows because we can examine the component of line-of-sight velocities normal to the disk. The target BHB sightlines lie at Galactic latitudes $b>60^{\circ}$, with the exception of three from the initial pilot study at $b=51.4^{\circ},55.2^{\circ},57.5^{\circ}$.

Finally, we defined a uniform sampling region in projected $x$-$y$ space across the disk to allow for comparison of extraplanar gas properties at different heights. Since the disk area spanned by BHBs at $b>60^{\circ}$ increases with distance, spatially uniform sampling requires that the radius of this cylinder is limited by the number of available sightlines at small heights above the disk. To achieve this uniform sampling, we chose sightlines which lie within a cylinder 2 kpc in radius, centered on the Sun and extending 15 kpc into the halo (Figure \ref{fig:keckoverview}).

\subsection{Coordinate System}
\label{sec:coordinatesystem}
\begin{figure}
    \centering
    \includegraphics[width=0.48\textwidth]{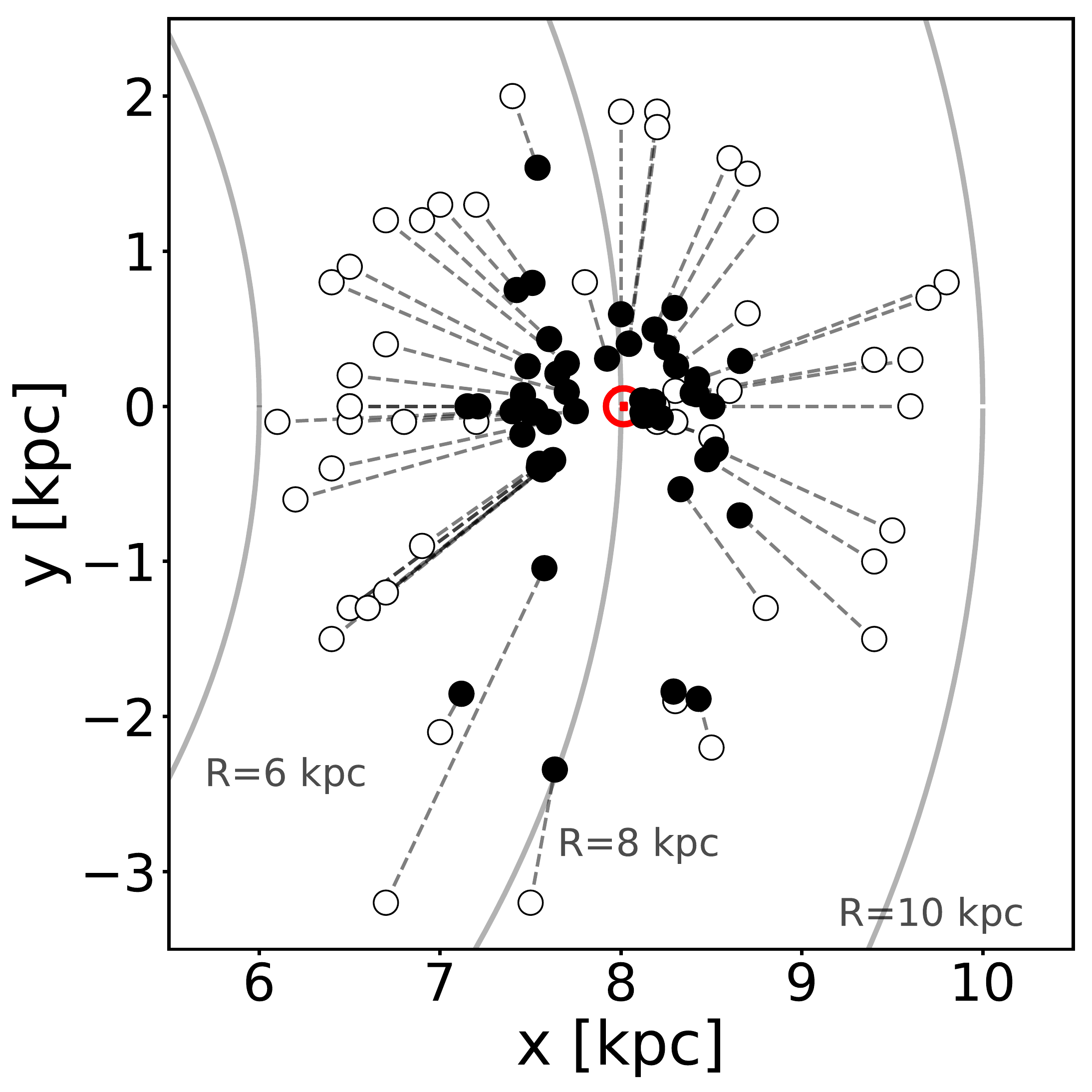}
    \caption{{\sc Effect of height assumptions on gas coordinates.} An illustration of the change in $x$-$y$ coordinates for hypothetical gas along each sightline when we assume two different distances to the gas. Open circles mark the coordinates of the upper limit on distance set by the background star. Filled circles, connected by a dashed line, show the location of gas along the same sightline assuming a height of z=3 kpc. The red solar symbol marks the position of the Sun, and lines of constant distance from Galactic center are shown in grey.}
    \label{fig:xycloudcoords}
\end{figure}

BHB distance measurements allow us to adopt 3-dimensional coordinates, rather than sky coordinates, to map the distribution of gas. A 3-D coordinate system is useful for studying flows in the frame of the Milky Way and is typical for Milky Way dynamical measurements, but is not often used in absorption studies because distance measurements are rarely available or are too imprecise. Throughout this paper we use a Cartesian coordinate system in our discussion of the position, scale and kinematics of gas near the disk. The $x$-axis extends from Galactic center in the direction of the Sun, $y$ is perpendicular to $x$ in the plane of the disk, and $z$ is perpendicular to the disk such that the system is right-handed, with positive $z$ towards the north Galactic pole. In this reference frame, the Galactic center is at the origin and the Sun lies at ($x$,$y$,$z$) = (8,0,0) kpc. Figure \ref{fig:keckoverview} shows an edge-on view of the Milky Way in the $x$-$z$ plane, and Figure \ref{fig:xycloudcoords} shows a face-on view of the disk with BHB coordinates projected onto the the $x$-$y$ plane.

We note that the BHB distance measurements represent an upper limit on the distance to any gas detected along the line of sight. Therefore any distances inferred from spatial maps of gas absorption are upper limits. To illustrate this point, Figure \ref{fig:xycloudcoords} provides a visual comparison of the $x$-$y$ coordinates of gas along each sightline assuming two different distances. Coordinates of the upper limit on distance set by the background star are marked by open circles. These are connected by a dashed line to filled circles marking the coordinates of gas along the same line of sight, assuming all gas lies at a height of $z=3$ kpc (the motivation for making such an assumption is discussed in \S\ref{sec:columndensity}). Although the disk area covered by this survey changes significantly, the results we present remain qualitatively the same regardless of the distance assumed. For that reason, we show only the coordinates corresponding to the background source in all subsequent plots unless otherwise noted.

\begin{figure*}
    \includegraphics[width=1.\textwidth]{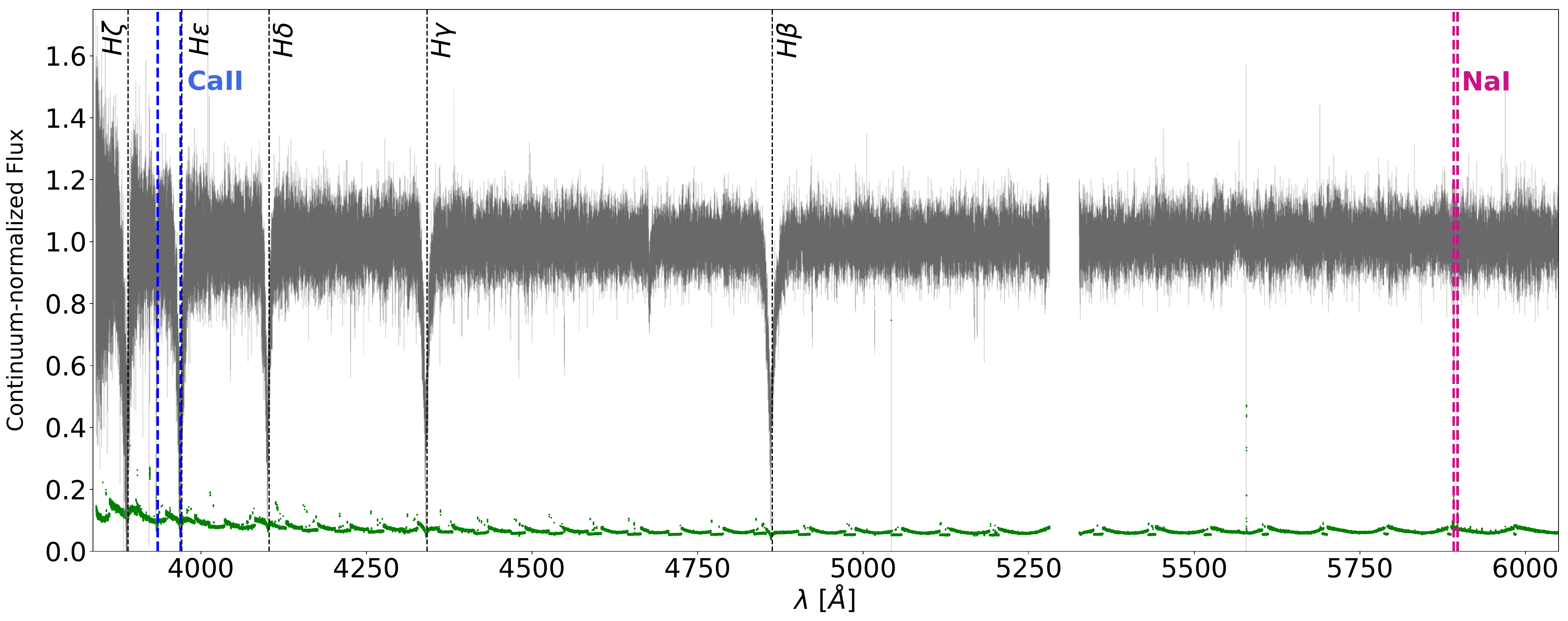}
    \caption{{\sc Spectrum of BHB used as background source.} Keck HIRES spectrum of J1217+2104, a typical metal-poor BHB used as a background source. Flux measurement error is shown in green. Vertical lines mark stellar Balmer absorption features as well as the CaII ($\lambda\lambda$ 3934.8, 3969.6 \AA) and NaI ($\lambda\lambda$ 5891.6, 5897.6 \AA) absorption doublets used in this work. Note that the CaII 3969.6 line falls within the broad $\rm H\epsilon$ feature, and that the continuum was adjusted to account for this feature before fitting the CaII doublet. A CCD chip gap falls at 5282-5326 \AA. The jagged profile of the flux error is due to the blaze function of each order, which results in more flux at the center of the order than at the edges.\\}
    \label{fig:spectrum}
\end{figure*}

\begin{figure*}[hp!t]
    \centering
    \includegraphics[width=0.9\textwidth]{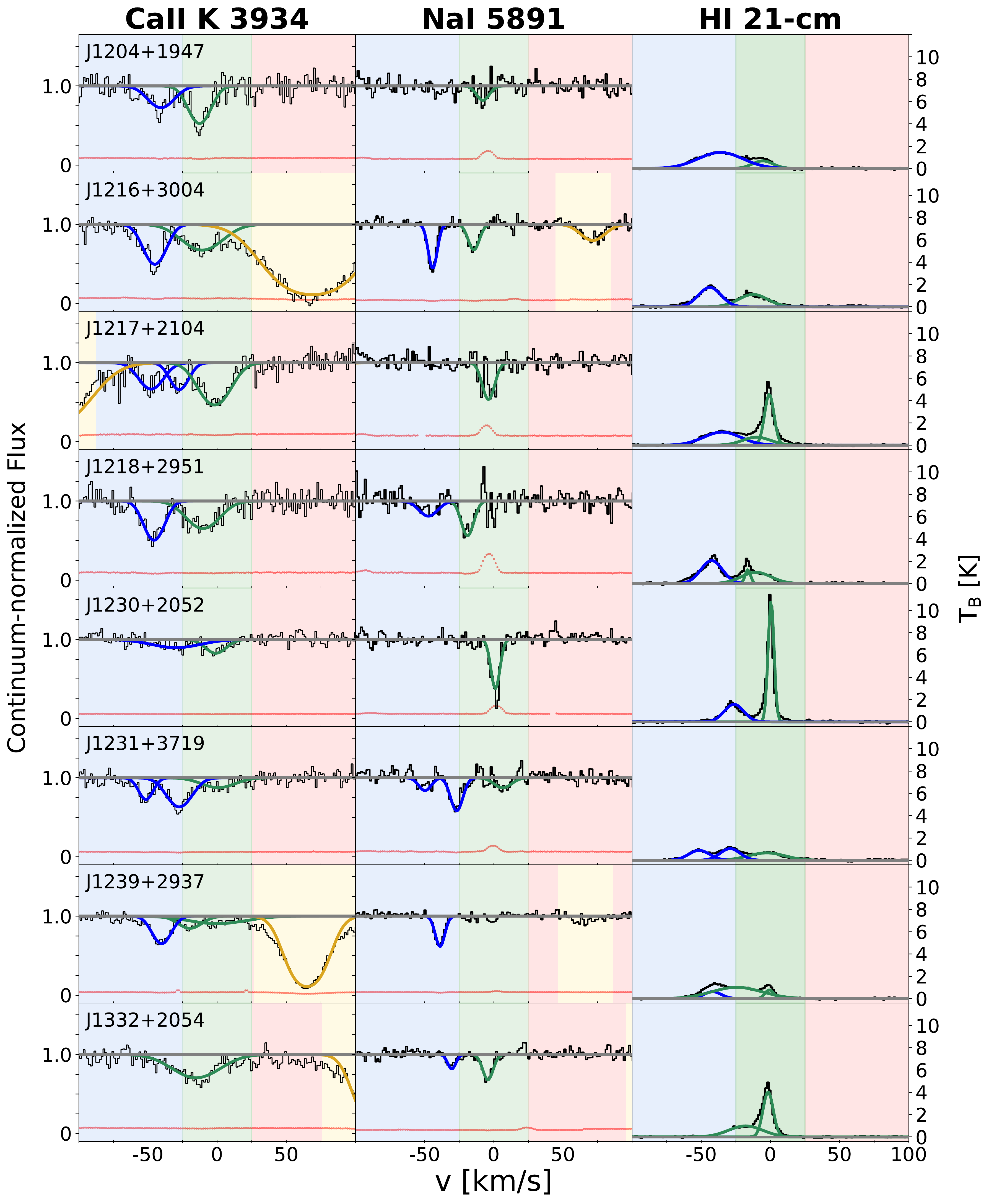}
    \caption{{\sc Detailed view of absorption features.} Examples of the CaII and NaI absorption features marked in Figure \ref{fig:spectrum} for several BHB sightlines (columns 1 \& 2), and HI 21-cm brightness temperatures measured nearest those sightlines (column 3) from the Effelsberg-Bonn HI Survey (\citealt{winkel2016}). Flux measurement errors for CaII and NaI are shown in red. The velocity range and absorption features associated with ISM gas ($|v|$ $<$ 25 km/s) are shaded green; otherwise positive velocities are shaded red and negative velocities are shaded blue. Regions near the stellar radial velocity and any associated absorption are highlighted in yellow if they fall within the window of the plot. While the spectra show stellar CaII absorption in some cases, they rarely show stellar NaI absorption. Almost all absorption detected outside the ISM velocity range is blueshifted, indicating a gaseous component moving towards the disk at approximately $-75$ to $-25$ km/s. The full set of line profile fits is included in Appendix \ref{appendix:linefits}.}
    \label{fig:abslineheights}
\end{figure*}

\subsection{Keck HIRES Spectra}
\label{sec:spectra}
At optical wavelengths, the HIRES spectrometer \citep{vogt1994} at the W.M. Keck Observatory can detect relatively weak absorption features of neutral and low-ionization species like CaII H+K and NaI D, which trace cool gas ($T\lesssim10^4 K$) \citep{munch1961,spitzer1956}. Figure \ref{fig:spectrum} is an example of a typical BHB spectrum and shows where the CaII and NaI absorption features fall with respect to the broad stellar Balmer lines. Columns 1 \& 2 of Figure \ref{fig:abslineheights} show a more detailed view of the CaII K and NaI D$_2$ absorption features for selected sightlines. We obtained HIRES spectra for 56 out of 83 total stars which met the criteria outlined above. Two spectra with $\rm S/N < 6$ were ultimately excluded from the sample because signal-to-noise was too low for reliable line profile fitting.

Observations were made over two runs in March 2016 and February 2017, and the instrument setup was chosen to optimize measurement of the CaII H+K ($\lambda\lambda$ 3934.8, 3969.6 \AA) and NaI D ($\lambda\lambda$ 5891.6, 5897.6 \AA) doublets. We took a minimum of two exposures for each sightline to allow for cosmic ray correction. Exposure times ranged from 300-1800 seconds, depending on the apparent magnitude of the target, resulting in S/N=6-46 per pixel with a median of 16.2 at wavelengths near the CaII doublet and S/N=9-60 per pixel with a median of 21.5 for the NaI doublet.
Typical wavelength coverage was $\sim$3800-8350 \AA, and when possible we adjusted wavelength coverage of each chip so that lines of interest would fall within the overlapping wavelength range detected by both chips.  We used the C5 decker, with the image rotator in vertical-angle mode. The C5 decker provides spectroscopic resolution of 1.3 km/s/pixel and $R=36$,$000$, which corresponds to a velocity FWHM of 8.3 km/s.

Raw data were reduced using the {\tt HIRedux}\footnote{http://www.ucolick.org/$\sim$xavier/HIRedux/} pipeline code included in {\tt XIDL}\footnote{http://www.ucolick.org/$\sim$xavier/IDL/index.html}, a data-reduction software package. We used the pipeline to combine multiple exposures for each sightline, calibrate spectra using flats and ThAr arcs, determine slit profiles for each order, perform sky subtraction, and remove cosmic rays. For a more detailed explanation of this procedure, we refer the reader to \cite{omeara2015}.

Automated flux calibration of HIRES spectra is difficult and unreliable, so each echelle order was examined by eye for continuum fitting. A Legendre polynomial was fit to each order using the AstroPy-affiliated
{\tt linetools} package\footnote{https://github.com/linetools/linetools}, an open-source code for analysis of 1D spectra. After automated calibration and co-adding of all orders was complete, the continuum was again examined and adjusted by hand by placing or moving anchor points in absorption-free regions. The 3934 \AA \, CaII K line often coincided with broad Balmer features in the stellar spectrum, in which case the continuum was adjusted to fit out these features and ensure that they did not affect column density measurements during analysis.

Finally, we corrected for cosmic rays, hot pixels, and gaps in the data which had not been removed by the {\tt HIRedux}  \ts pipeline. \ts Each \ts affected \ts pixel \ts was \ts assigned \ts a \newpage \noindent new value equal to the mean of the nearest unaffected pixels on each side.
\\

\section{Analysis}
\label{sec:analysis}
\subsection{Line profile fitting}
\label{sec:linefitting}
For each sightline, we used CaII and NaI absorption features, along with HI 21cm emission measurements from the Effelsberg-Bonn HI Survey (EBHIS), to identify distinct components of gas in velocity space and fit line profiles to them. The features and line profiles discussed in this section are shown in Figure \ref{fig:abslineheights} for  CaII K ($\lambda$ 3934 \AA), NaI D$_2$ ($\lambda$ 5891 \AA), and HI 21-cm along several sightlines.

\subsubsection{Ca II and Na I absorption}
\label{sec:absorption}
We fit line profiles to the CaII H+K ($\lambda\lambda$ 3934.8, 3969.6 \AA) and NaI D ($\lambda\lambda$ 5891.6, 5897.6 \AA) doublets in order to measure gas velocity $v$, column density $N$, and Doppler b parameter $b_{\rm D}$. The Doppler parameter is a measure of line width, and reflects line broadening caused by both thermal and non-thermal processes. The wavelength range spanned by each absorption feature and initial guesses for the fit parameters were determined by hand using the {\tt PyIGM IGMGUESSES} GUI\footnote{https://github.com/pyigm/pyigm}. Individual inspection allowed us to use priors of narrow velocity width and mask regions of severe blending when fitting multiple velocity components, which was crucial for obtaining good fits since all absorption features were fit independently of one another. These initial guesses were then passed to a fitting routine that uses the Markov Chain Monte Carlo (MCMC) technique as implemented in the Python package {\tt emcee} \citep{emcee} to calculate posterior probability distribution functions (PPDFs) for each model parameter.  We assume that the logarithm of the likelihood function is equal to the distribution of $\chi^2/2$, adopting uniform priors over the parameter ranges $0.5 < b_{\rm D} < 75$, $\rm 8 < logN < 13$. We use Markov Chains produced by 50 walkers taking 500 steps, with a burn-in of 150 steps. We report the median and $\pm 34$th percentiles of the marginalized PPDFs as the parameter values and uncertainty intervals for each parameter, respectively.

Once the heliocentric velocities were determined by the MCMC fitting routine, they were corrected for motion relative to the Local Standard of Rest (LSR) along each sightline, so that gas which is moving at the LSR has $v_{\rm LSR}=0$ km/s. All velocities in the following discussion are defined relative to the LSR in this way unless otherwise noted. We do not assume rotation for extraplanar gas, as any effects from rotation or halo lag are orthogonal to the inflow velocities we observe. Because our sightlines are at high Galactic latitudes ($b>60^{\circ}$, except three at $b=51.4^{\circ},55.2^{\circ},57.5^{\circ}$), Galactic rotation effects are not significant for line-of-sight velocities in this sample \citep{wakker1991}. For the 48/54 sightlines at $b > 70^{\circ}$, Galactic rotation introduces $< 5\%$ error in velocity, and for the remaining six sightlines at $50^{\circ} < b < 70^{\circ}$ the velocity error is $< 12\%$. The inflow velocities we measure vary by considerably more than 12\%, which means that the trends we observe are not due to rotation effects. Lastly, if no absorption was detected, noise in the spectra near the wavelength of the line was used to calculate $2\sigma$ upper limits on equivalent width and column density. Across all sightlines, the column density detection limit had a median logN of 10.97 \cm for CaII and 10.58 \cm for NaI.

\subsubsection{HI 21-cm emission}
\label{sec:HI}
For purposes of comparison to the more complete observations which have been made in HI 21-cm emission, we obtained data from the Effelsberg-Bonn EBHIS HI survey \citep{winkel2016} and analyzed HI emission for positions closest to each BHB sightline. The positions do not precisely match our BHB sightlines because of differing spatial resolutions; the effective beam size for EBHIS is 10.8\arcmin \ compared to the 1.0\arcsec \ slit used for HIRES observations \citep{vogt1994}. The spectral resolution of EBHIS is $\sim$1 km/s. HI brightness temperature profiles were inspected by eye to obtain initial guesses for line profile parameters and determine the number of distinct emission features in velocity space. These components were fit simultaneously to a Gaussian mixture model, which determined the best fit with the Levenberg-Marquardt damped least-squares algorithm using the SciPy {\tt curve\_fit} optimizer \citep{scipy}. Note that since HI is detected in emission rather than in absorption along BHB sightlines, the upper limits on distance provided by background sources for CaII and NaI do not apply to HI.
\\

\subsection{Isolating Absorption Signatures of the Galactic Fountain}
The distances to BHBs set upper limits on the distance to any intervening gas clouds along the line of sight. Because absorption could originate from gas anywhere along the sightline, extraplanar flows can't be distinguished from the Milky Way ISM concentrated in the Galactic disk using only the upper limits on distance available to us. However, extraplanar gas flows do exhibit kinematic signatures that allow us to differentiate them from Milky Way ISM gas.

In order to isolate absorption features associated with Galactic inflows/outflows and examine their properties, we separate absorption lines into two distinct velocity components: `ISM' (velocities consistent with the Milky Way's interstellar medium, $|v|<25$ km/s) and `IV' (Intermediate Velocity gas, $|v|>25$ km/s). The first is a component we associate with the Milky Way ISM; this is gas moving within $\pm$25 km/s of the local standard of rest. Any gas outside this velocity range is not likely to be part of a coherent rotating disk, and may be associated with extraplanar gas flows. This so-called IV gas component is often defined as $35<|v|<90$ km/s, but this range is somewhat arbitrary and the cutoffs are often adjusted based on the gas velocities of features being considered \citep{vanwoerden2005}. In this work we do not detect any gas at $|v|>75$ km/s, and find that adopting a cutoff at $|v|=25$ km/s for ISM gas best reveals the trends in these data.

It should be noted that interpretations of IV gas measurements are seldom clear-cut. At high Galactic latitudes, typical Milky Way disk gas should have low line-of sight velocities, since the bulk of its motion is in the transverse direction. However, any extraplanar gas moving at $|v|<25$ km/s cannot be easily disentangled from disk gas \citep{boettcher2017,zheng2017b}.
\\

\section{Results}
\label{sec:results}

\begin{figure}
    \centering
    \vspace{15pt}
    \includegraphics[width=0.48\textwidth]{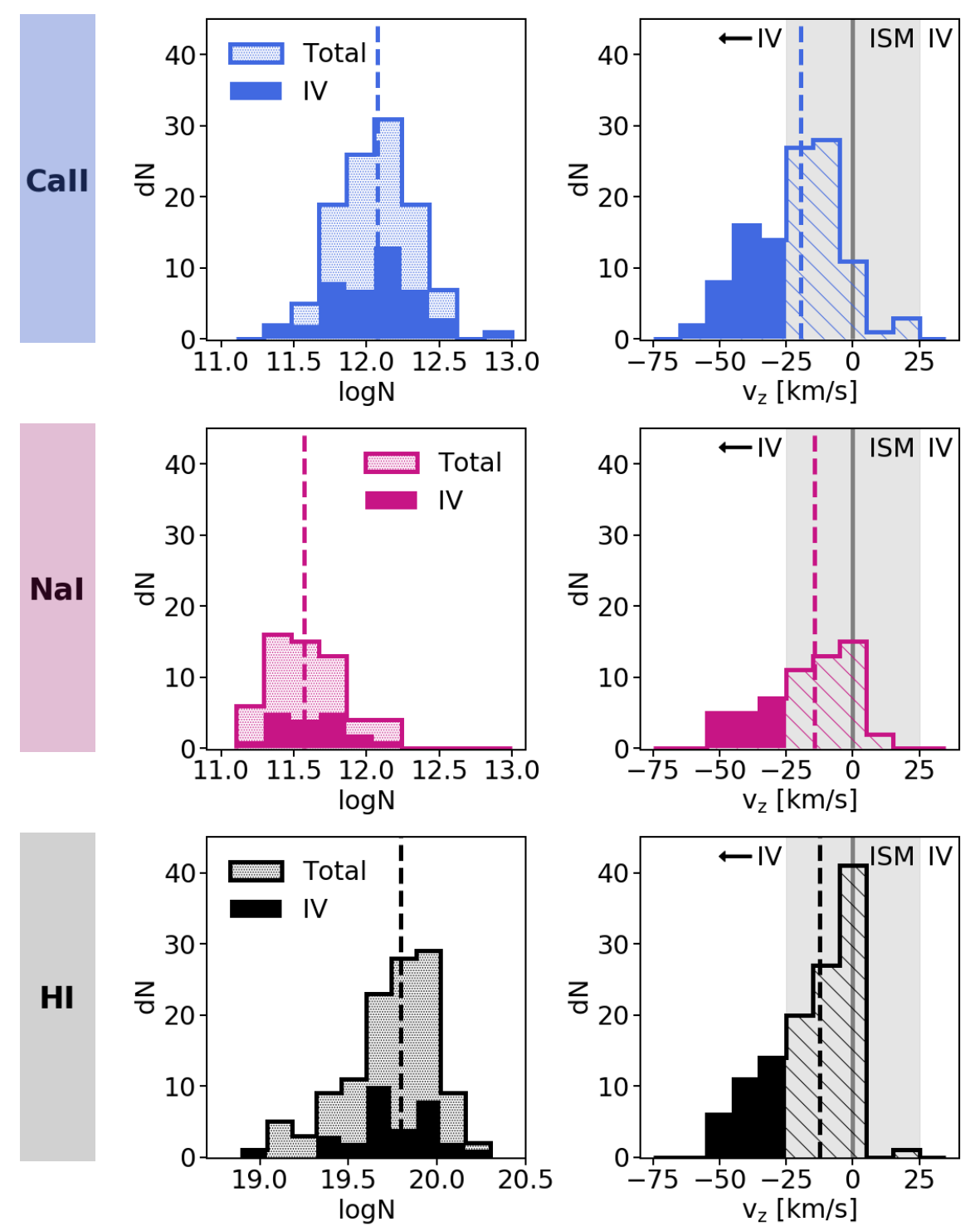}
    \caption{{\sc Distribution of column density \& velocity.} Measured column densities (left column) and the $z$-component of measured radial velocities (right column) for CaII, NaI, and HI. Dashed lines mark the median value for each distribution, and in the right column $v_{\rm LSR} = 0$ is indicated with a grey line. All ions are dominated by Galactic signals at $|v|<25$ km/s (hatched fill), but a significant IV component at $v<-25$ km/s can be clearly seen (solid fill), indicating inflowing gas at the disk-halo interface.\\}
    \label{fig:vhist}
\end{figure}

Table \ref{table:mcmclist} lists best-fit values for column density $N$, radial velocity $v$, and Doppler b parameter $b_{\rm D}$, which we combine with spatial coordinates from the SEGUE catalog to provide a characterization of IV gas in the halo. Figure \ref{fig:vhist} shows the distribution of the CaII, NaI, and HI parameter values fit using the methods described in \S\ref{sec:linefitting}.

We have tabulated results for gas detections at all velocities, but limit discussion in the body of this paper to the IV gas component. The results and figures for the Milky Way ISM component ($|v|<25$ km/s) are presented in Appendix \ref{appendix:MW}. This low-velocity gas behaves largely as expected for typical disk material, but we examine some noteworthy features of this component as they relate to IV gas.

\begin{figure*}[p]
    \centering
    \includegraphics[width=\textwidth]{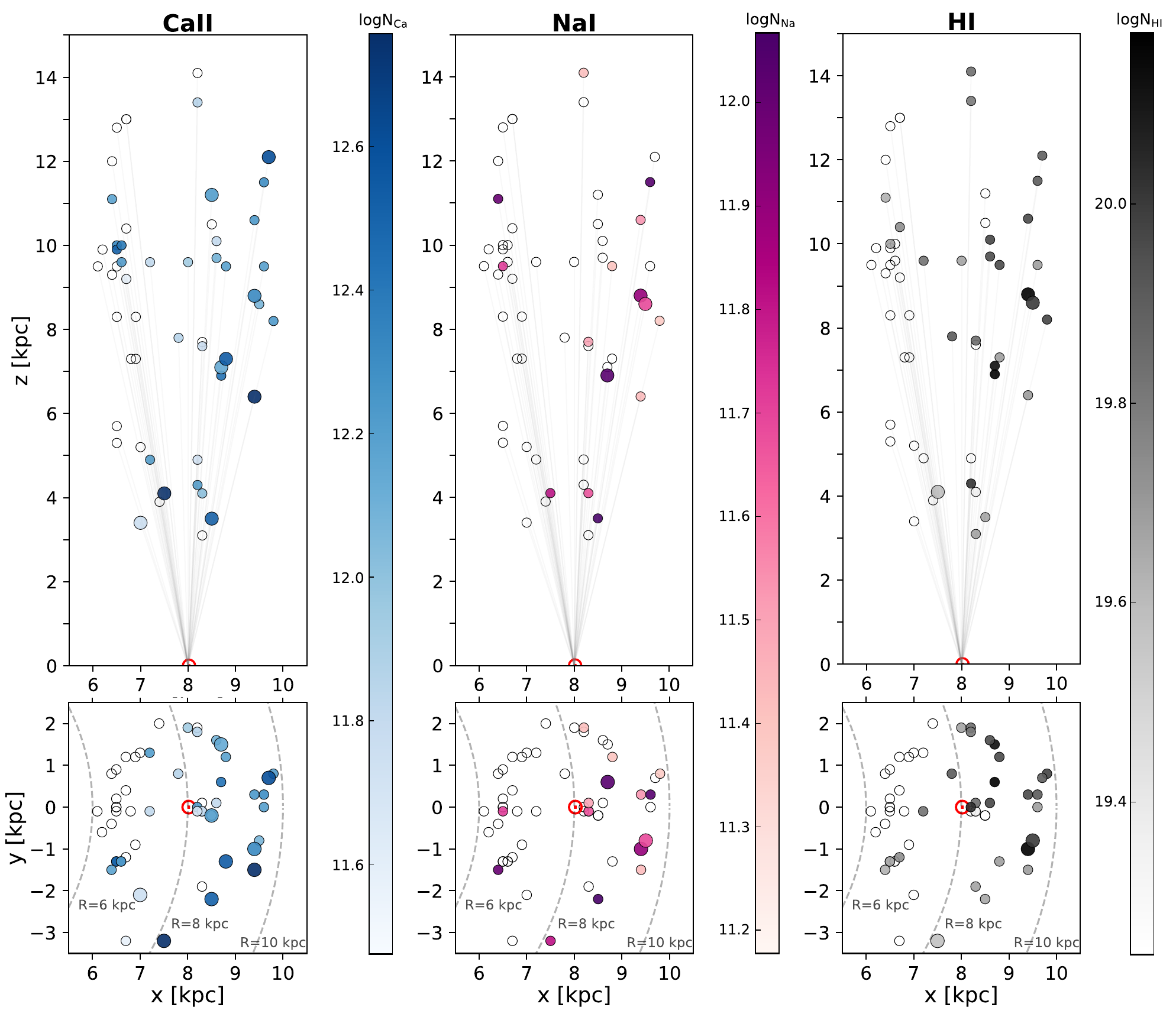}
    \caption{{\sc Column density measurements along BHB sightlines.} Column densities for the IV component (Intermediate Velocity gas, $v<-25$ km/s). Markers correspond to background BHB sightlines projected onto the $x$-$y$ and $x$-$z$ planes of the Milky Way disk. The coordinates for each BHB are plotted in physical kpc units in the plane of the Milky Way; thus any inferred distance to gas absorption along a sightline is an upper limit. Color indicates column density measurements for CaII and NaI in absorption and HI in 21-cm emission. Larger markers indicate sightlines along which more than one component was detected, and their colors represent the total column density of those components. Empty circles denote nondetections. The Sun is marked by a red solar symbol, and black dashes denote lines of constant radius from Galactic center. IV column densities do not uniformly increase with z or BHB distance, implying that most of the gas is closer than the nearest background source at $z \lesssim 3.1$ kpc. }
    \label{fig:IVcol}
\end{figure*}

\begin{figure}
    \centering
    \vspace{10pt}
    \includegraphics[width=0.48\textwidth]{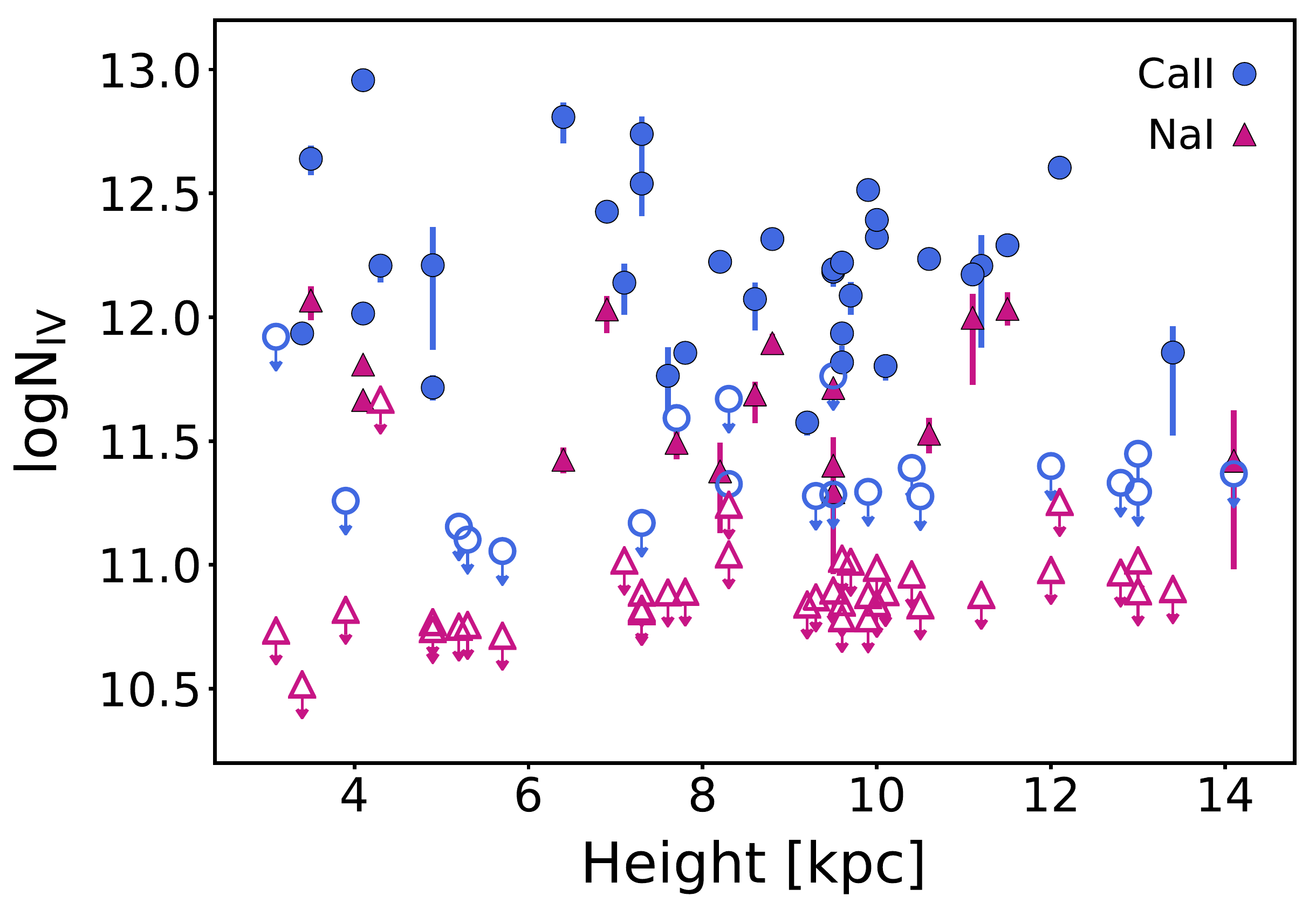}
    \caption{{\sc CaII and NaI lie close to the disk.} The total column density of IV gas detected along BHB sightlines, which have background sources at varying heights above the disk. CaII measurements are marked by blue circles, and NaI measurements by magenta triangles. For sightlines along which no gas was detected, $2\sigma$ upper limits on column density are shown with unfilled markers. There is no correlation between the distance probed by a sightline and the total column density of gas, suggesting that most of the gas detected lies closer than the nearest background source at $z \lesssim 3.1$ kpc. \\}
    \label{fig:NvsZ}
\end{figure}

\begin{figure*}[!htb]
    \centering
    \includegraphics[width=0.85\textwidth]{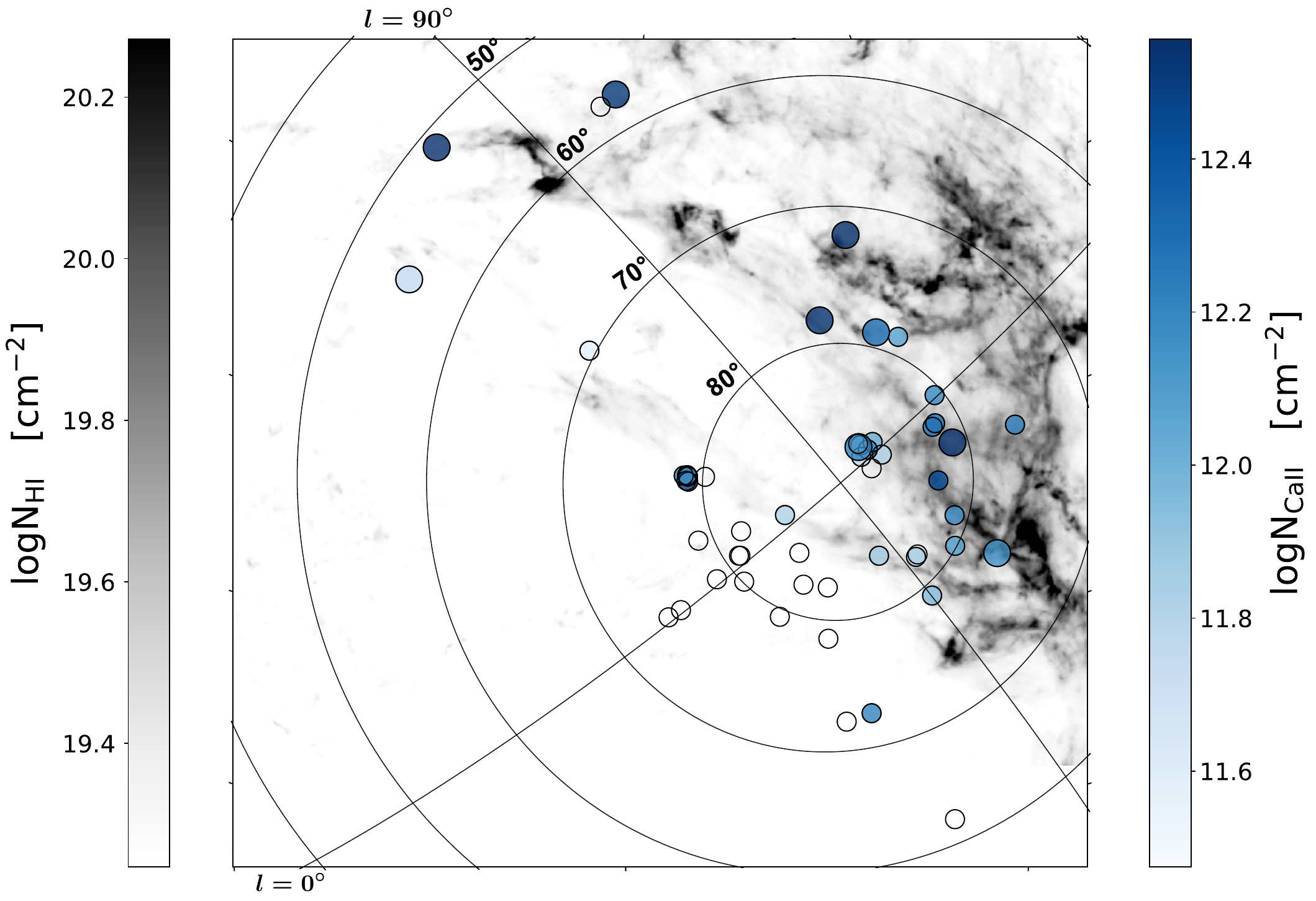}
    \caption{{\sc CaII column density comparison to HI 21-cm map.} A map of HI column density at $-60<v<-25$ km/s from the LAB survey in grey \citep{labsurvey}, overplotted with CaII IV gas column densities from this work in blue. Each marker represents a BHB sightline, colored by the total IV gas column density along that sightline. Open circles denote nondetections. Larger markers indicate sightlines with multiple distinct IV absorption components, and are colored by the total IV column density of those components. CaII detections correlate with cool gas near the edge of two large HI complexes known as the IV Arch and the IV Spur.\\}
    \label{fig:WCS_N}
\end{figure*}

\subsection{Column Density}
\label{sec:columndensity}
Absorption line measurements reveal how gas column densities vary across the face of the Milky Way's disk. When these measurements are combined with distances to background BHBs, we can also see how column densities vary with height above the disk, providing a 3-dimensional picture of the absorbing gas. Figure \ref{fig:IVcol} shows total column densities for the IV gas component along each sightline, from both an edge-on view (top panel) and projected onto the plane of the Milky Way's disk (bottom panel). The markers are located at the coordinates of the background BHB; in other words, they reflect the upper limit on distance to the absorbing CaII and NaI gas. Results remain qualitatively the same if smaller distances are assumed (see Figure \ref{fig:xycloudcoords} for an example of the effect of distance assumptions on gas cloud coordinates). Colors indicate the column density of IV gas, with darker colors corresponding to greater column densities and empty circles denoting nondetections. Larger markers represent sightlines with multiple distinct IV gas components which have each been fit separately, and their colors represent the summed column density of those components.

In the $x$-$y$ plane, most detections of IV gas are grouped together along sightlines outside the solar circle, while very little IV gas is found in the direction towards the Galactic center. This pattern is consistent across the CaII, NaI, and HI tracers. Although CaII and HI detections appear spatially coincident, we find no significant correlation between CaII and HI column densities as one might expect (Pearson $r = -0.04$, $p = 0.86$). We speculate that this may be an effect of the 10.8\arcmin \ beam size of the EBHIS HI data, which smears out signal over a large area relative to the 1\arcsec \ slit width for HIRES. Alternatively, the absence of any correlation in CaII-HI column density could be real if the HI is associated with gas in a more highly ionized phase.

Additionally, there is no significant correlation between column densities and the height probed by a sightline. This absence of any trend between total column density along a sightline and the height of the background source above the disk can be seen more clearly in Figure \ref{fig:NvsZ}. If the gas were distributed uniformly or exhibited a smooth dropoff with height, we would expect to observe higher column densities along sightlines to more distant background sources. We find no evidence of such correlation (Pearson $r = -0.09$, $p = 0.64$), an indication that most of the observed gas lies at a distance closer than $z = 3.1$ kpc, the distance to the nearest background source. Further support for this picture can be seen in Figure \ref{fig:WCS_N}, in which our CaII IV gas column density measurements are overlaid on a map of HI column density from the LAB survey \citep{labsurvey}. These CaII detections visibly coincide with the edge of two large HI complexes called the IV Arch and the IV Spur, which extend across much of the northern Galactic sky, and are estimated to sit above the Galactic disk at distances of 0.5-3 kpc (\citealt{wakker2001}; \citealt{kuntzdanly1996}; \citealt{smoker2011}).
\\

\subsection{Kinematics}
\label{sec:kinematics}
\begin{figure*}[p]
    \centering
    \includegraphics[width=\textwidth]{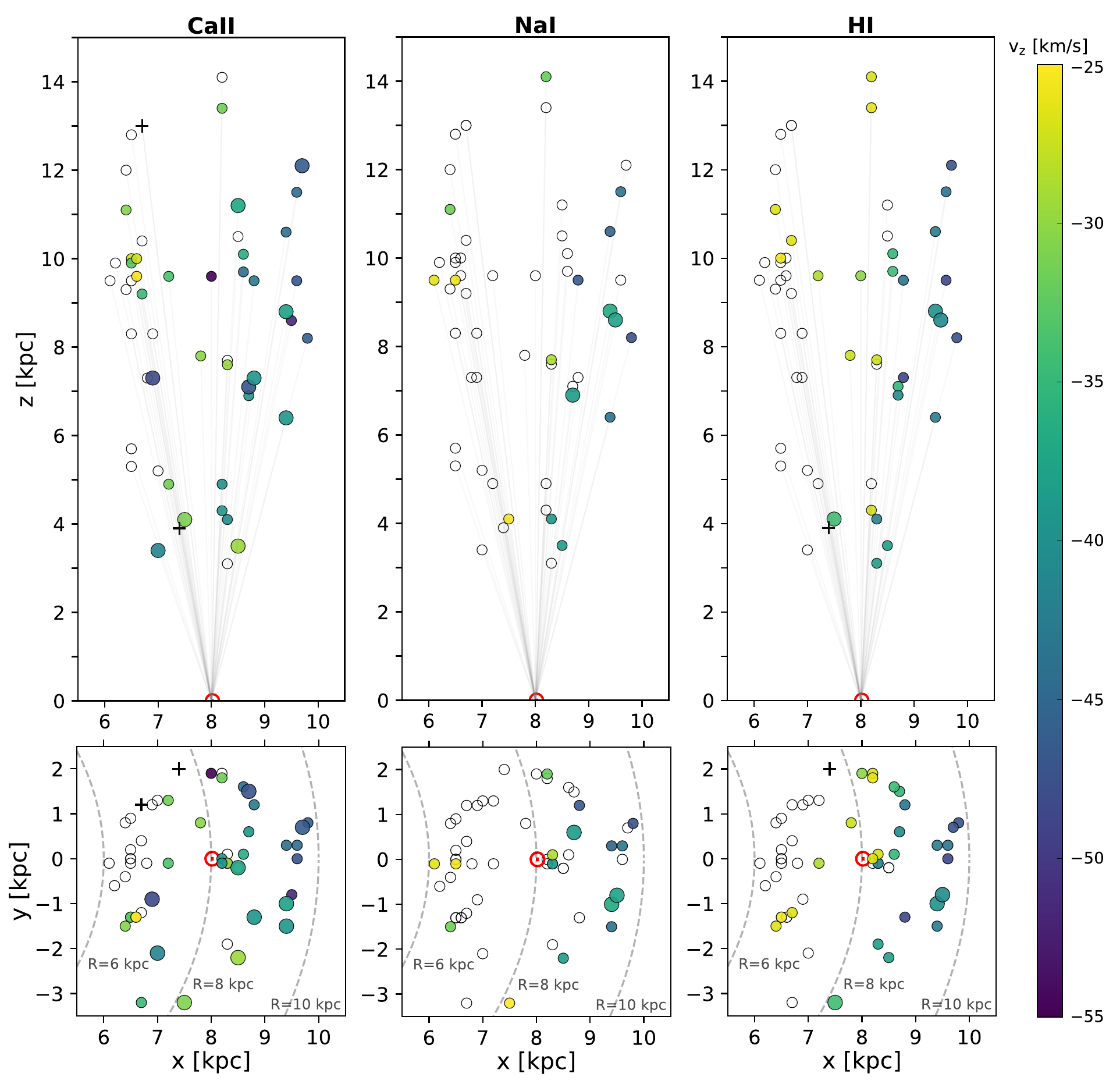}
    \caption{{\sc Infall velocity measurements along BHB sightlines.} Infall velocities for the IV component (Intermediate Velocity gas, $v<-25$ km/s). Markers correspond to background BHB sightlines projected onto the $x$-$z$ ({\it top}) and $x$-$y$ ({\it bottom}) planes of the Milky Way disk. The coordinates for each BHB are plotted in physical kpc units in the plane of the Milky Way; thus any inferred distance to gas absorption along a sightline is an upper limit. Color indicates the component of radial velocity measurements perpendicular to the disk ($v_{\rm LSR}$ sin$b$) for CaII and NaI in absorption and HI in 21-cm emission. Larger markers indicate sightlines along which more than one component was detected, and their colors represent the mean velocity of those components weighted by column density. IV gas is only detected at negative velocities, with the exception of one absorber just beyond the IV cutoff in CaII and HI, and another at $v=40.8$ km/s in CaII. These absorbers are marked by a `+' symbol. Empty circles denote nondetections. The Sun is marked by a red solar symbol, and black dashes denote lines of constant radius from Galactic center. The infalling gas exhibits a clear velocity gradient, especially in CaII. By comparison, the `ISM' gas shows no such trend (see Appendix \ref{appendix:MW}).}
    \label{fig:IVvel}
\end{figure*}

Figure \ref{fig:IVvel} maps infall velocities for the IV gas component, again from both an edge-on view (top panel) and projected onto the plane of the Milky Way's disk (bottom panel). Colors represent the z-component of measured radial velocities ($v_z = v\,$sin$b$) in order to show motion directly towards/away from the disk. This survey has been designed to optimize constraints on the vertical component of gas velocities, so transverse velocities are not shown because they are not well-constrained by our experiment. However, results do not differ significantly if the observed radial velocity is used instead. Larger markers indicate sightlines with multiple distinct IV absorption features, and their colors represent the logN-weighted mean velocity of those components. A `+' symbol marks the location of two positive-velocity absorbers just beyond the 25 km/s IV cutoff in CaII and HI for J1223+0002a, and another in CaII at 40.8 km/s for J1258+1939. These absorption features are shown in Appendix \ref{appendix:linefits} and the parameters of their line profile fits are listed in Table \ref{table:mcmclist}. The distance constraints on our sightlines allow us to look at gas velocities in a unique way by projecting measured velocities on the $x$-$y$ plane of the disk. As in the corresponding column density plot (Figure \ref{fig:IVcol}), the location of each point reflects the upper limit on distance to the absorber, which lies somewhere between the background star and the Sun. The location of the Sun is indicated with a red symbol and dashed lines indicate lines of constant radius from the Galactic center.
 
\begin{figure}
    \centering
    \includegraphics[width=0.48\textwidth]{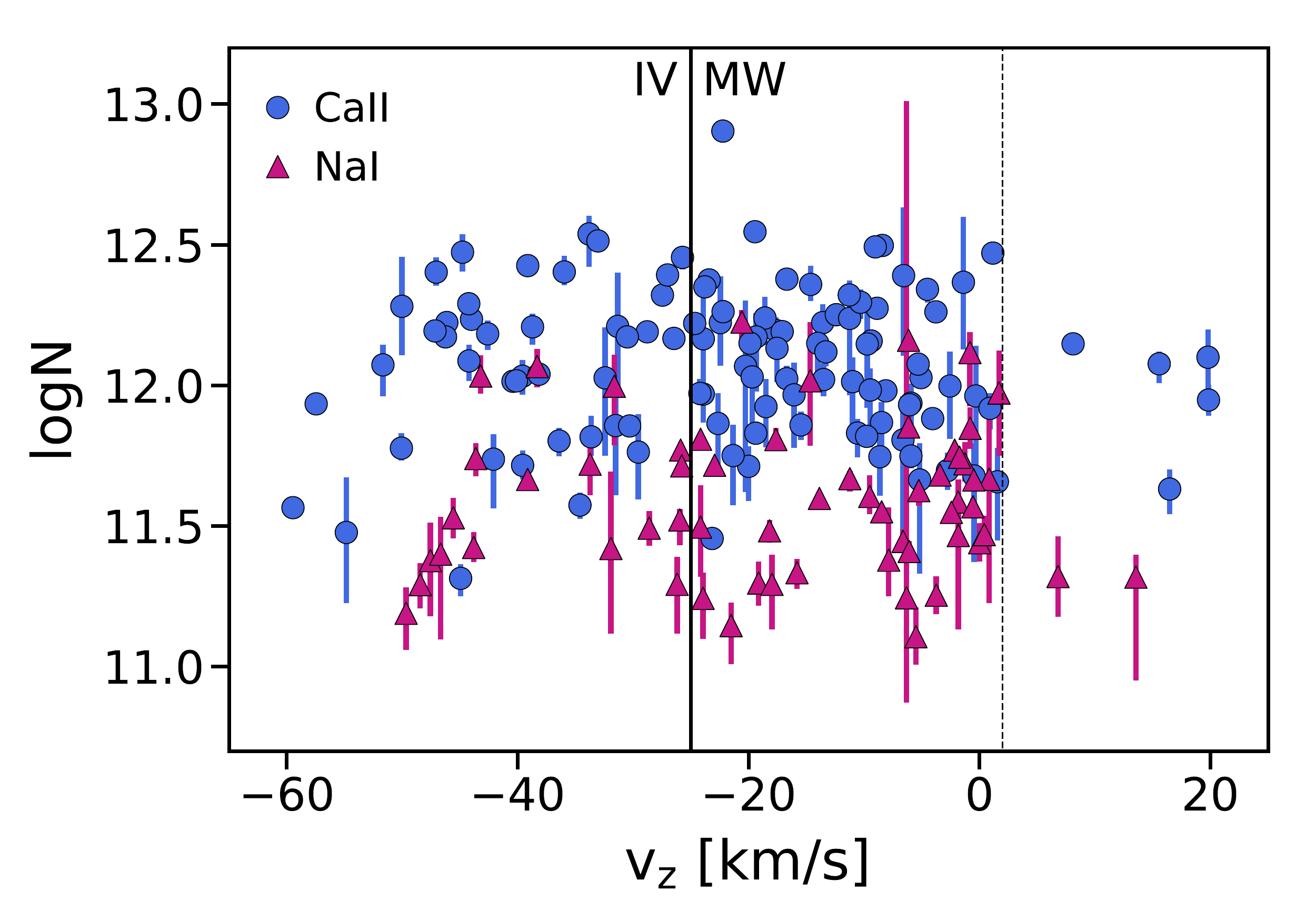}
    \caption{{\sc Column density \& velocity are uncorrelated.} Column density and $z$-component velocity for all distinct IV absorption components. CaII measurements are marked by blue circles and NaI by magenta triangles. There is no apparent relationship between the column density of the gas and its infall velocity.  Note the abrupt dropoff in CaII and NaI detections at $v_z \gtrsim 2$ km/s, marked by the vertical dotted line. This dearth of cool gas at positive velocities is also present in HI. \\} 
    \label{fig:Nvsv}
\end{figure}

The IV component we have isolated reveals a distinct pattern in vertical infall velocity. Virtually all of the IV gas within 13 kpc is at negative velocities, falling towards the disk. We detect only two IV CaII absorbers at positive velocities, and a single positive-velocity HI IV absorber just beyond the ISM velocity cutoff ($v = 25$ km/s). This is consistent with previous findings that gas in this HI complex, as well as the vast majority of IV gas at northern Galactic latitudes, is moving towards the disk \citep{kuntzdanly1996,albertdanly2004,richter2017}. We find no correlation between velocity and height above the disk; this provides further support for the picture suggested by our column density measurements in which IV gas sits mostly close to the disk at small heights (see \S\ref{sec:columndensity}). Furthermore, we find no correlation between velocity and column density (see Figure \ref{fig:Nvsv}).

\begin{figure}
    \centering
    \includegraphics[width=0.48\textwidth]{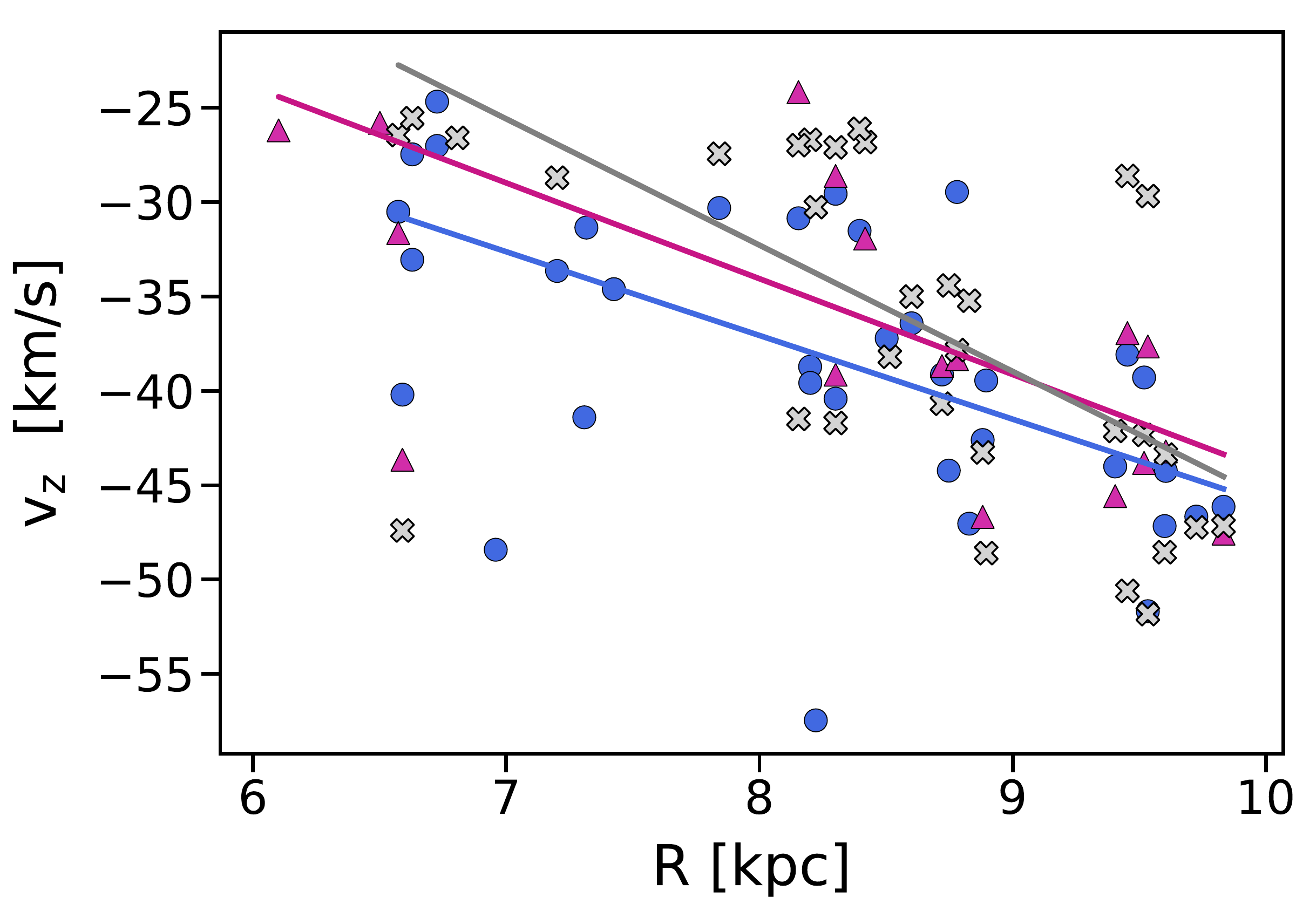}
    \caption{{\sc A velocity gradient across the disk.} IV gas velocities for CaII and NaI detections in absorption and HI 21-cm emission, showing the $z$-component of measured radial velocity as it varies with distance from Galactic center. For sightlines with more than one IV component, the logN-weighted velocity is shown. Lines show the least-squares fit to the data, revealing velocity gradients of $-6.0 \pm 1.5$ km/s/kpc in CaII, $-7.0 \pm 1.7$ km/s/kpc in NaI, and $-8.8 \pm 1.3$ km/s/kpc in HI. \\}
    \label{fig:veltrendfits}
\end{figure}

IV gas does exhibit a clear infall velocity gradient across the face of the disk which is oriented roughly along the radial axis of the Milky Way. Gas infall velocities at $R$ = 6-10 kpc increase in magnitude farther from the Galactic center ($-6.0$ $\pm$1.5 km/s/kpc in CaII, $-7.0$ $\pm$1.7 km/s/kpc in NaI, and $-8.8$ $\pm$1.3 km/s/kpc in HI). This velocity gradient is especially prominent for CaII and HI, as seen in the bottom panel of Figure \ref{fig:IVvel}. A similar trend appears in NaI, although there are fewer NaI detections overall. The gradient spans velocities from the IV cutoff at $-25$ km/s to $-75$ km/s, with faster-moving gas detected near the interior of the coincident HI IVCs mentioned in \S\ref{sec:columndensity}. In Figure \ref{fig:veltrendfits}, least-squares linear fits to this gradient for CaII, NaI, and HI show that this gradient is not explained by a geometric viewing effect, which would cause the $z$-component of observed radial velocities to be smaller for sightlines at lower Galactic latitudes, not larger.

\begin{figure*}[t]
    \centering
    \includegraphics[width=0.75\textwidth]{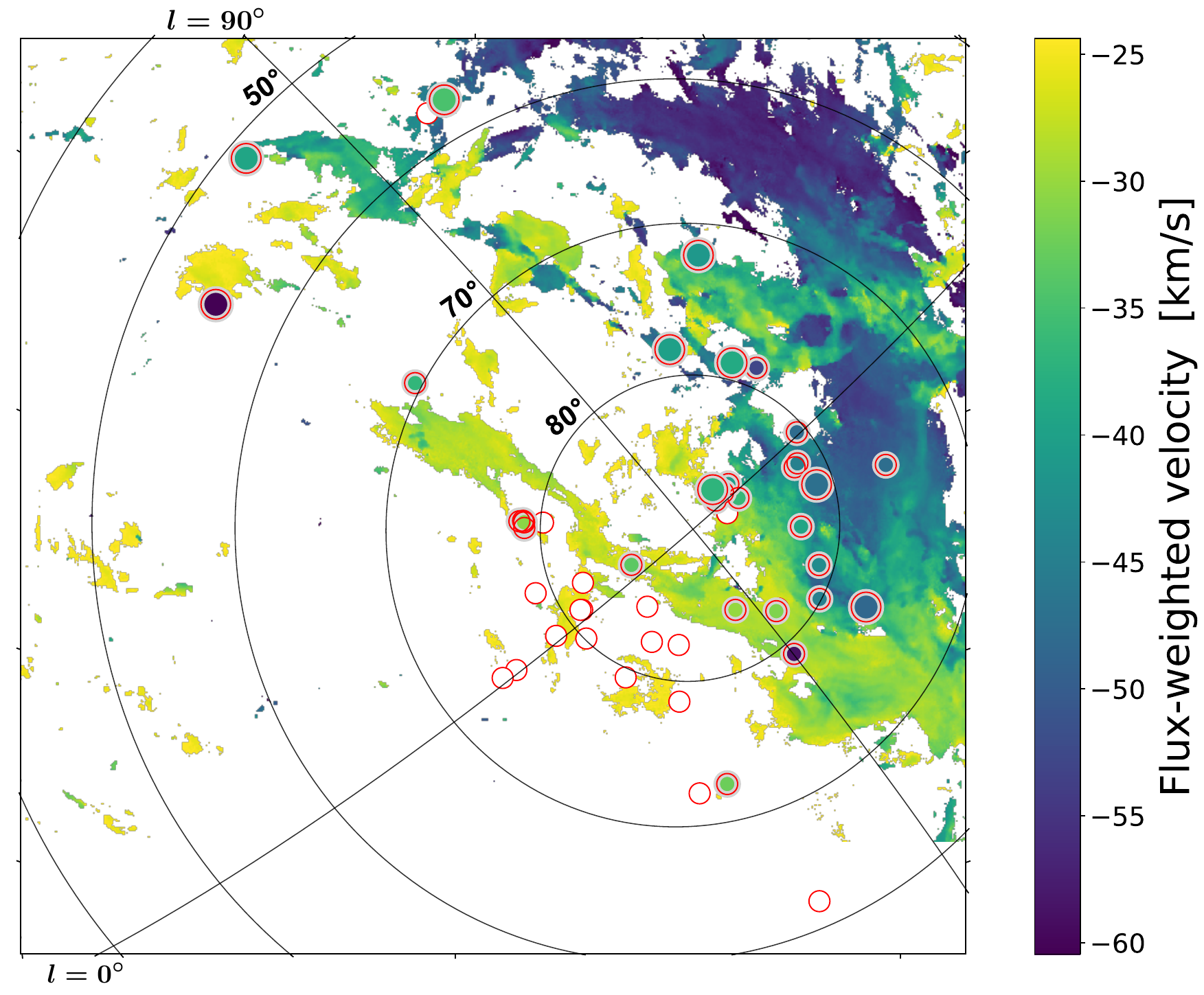}
    \caption{{\sc CaII velocity comparison to HI 21-cm map.} A map of flux-weighted HI velocities at $-60<v<-25$ km/s from the LAB survey \citep{labsurvey}, overplotted with column-density weighted infall velocities of CaII from this work. Each marker represents a BHB sightline, colored by the velocity of IV gas along that sightline. Open circles denote nondetections. Larger markers indicate sightlines with multiple IV absorption components at distinct velocities, and their color represents the mean velocity of those components weighted by column density. CaII detections correlate with the edge of two large HI complexes known as the IV Arch and the IV Spur. A velocity gradient can be seen in CaII, with faster infall velocities nearest the interior the cloud. \\}
    \label{fig:WCS_v}
\end{figure*}

The CaII velocity measurements are overlaid on a map of flux-weighted HI velocities from the LAB survey \citep{labsurvey} in Figure \ref{fig:WCS_v}. As mentioned above and shown in Figure \ref{fig:WCS_N}, CaII gas detections coincide with a large HI IVC complex, probing an area at the edge of the well-studied IV Arch and IV Spur. The locations of our CaII detections in the $x$-$y$ plane are spatially coincident with that HI structure and have similar velocity gradients. However, the CaII column densities and HI column densities are not correlated, which may be a result of mismatched spatial scales probed by emission and absorption observations, respectively. Such properties of HI and CaII may indicate that large coherent structures are accreting onto the Milky Way in a manner more complex than a single-slab accreting layer \citep{zheng2017a,forbes2018}. We explore various galactic fountain scenarios in our Discussion section (\S\ref{sec:galacticfountain}).

\subsection{Cloud Size \& Distribution}
\label{sec:cloudsizeanddist}

Given that the majority of detections are grouped together at $x\gtrsim 8$ kpc, it is likely that we are mapping the edge of a large structure $>3$ kpc in extent. Such an extended structure may be comprised of many smaller gas cloud fragments or filaments that effectively provide uniform coverage, as is the case for some HVCs \citep{putman2002,stanimirovic2008,hsu2011,wakker2001}. The substructure created by these so-called clouds is probably comprised of many interrelated clumps and wisps, rather than discrete structures with well-defined boundaries \citep{mccourt2018,fukui2018, marinacci2010, bekhti2007}.

Comparing CaII and NaI detections to existing HI maps provides further support for wispy substructures within clouds that extend beyond the coverage of this survey. Figure \ref{fig:WCS_N} shows HI emission from the LAB survey in grey, and such morphology is clearly visible \citep{labsurvey}. Our IV gas detections correlate strongly with the IV Arch and IV Spur HI complexes which extend across much of the northern Galactic sky and contain clumpy substructures with angular sizes of 10-20$^{\circ}$ \citep{kuntzdanly93}.

Distance estimates for the HI IV Arch and Spur generally fall within a range of 0.5-3 kpc (e.g. \citealt{wakker2001}; \citealt{kuntzdanly1996}; \citealt{smoker2011}), 
which is consistent with the distance constraint of $z \lesssim 3.1$ kpc we have determined for detected gas in this sample. If low-ionization Ca and Na can be tied to HI IVCs close to the disk, then measurements of HI gas provide additional spatial constraints. To illustrate this, Figure \ref{fig:xycloudcoords} maps the $x$-$y$ coordinates of gas assuming a cloud at z=3 kpc, compared to the $x$-$y$ coordinates corresponding to the upper limit on distance set by the background star. We assume a cloud height of z=3 kpc for the purposes of demonstration, since our detections place only an upper limit on this value. This figure highlights the significant effect that distance measurements can have on our ability to constrain substructure and cloud size. Connecting CaII and NaI gas with HI lying close to the disk is therefore a useful link that allows us to place tighter constraints on the scale of structures within the larger cloud probed by this survey.

\begin{figure*}[p!]
    \centering
    \includegraphics[width=0.95\textwidth]{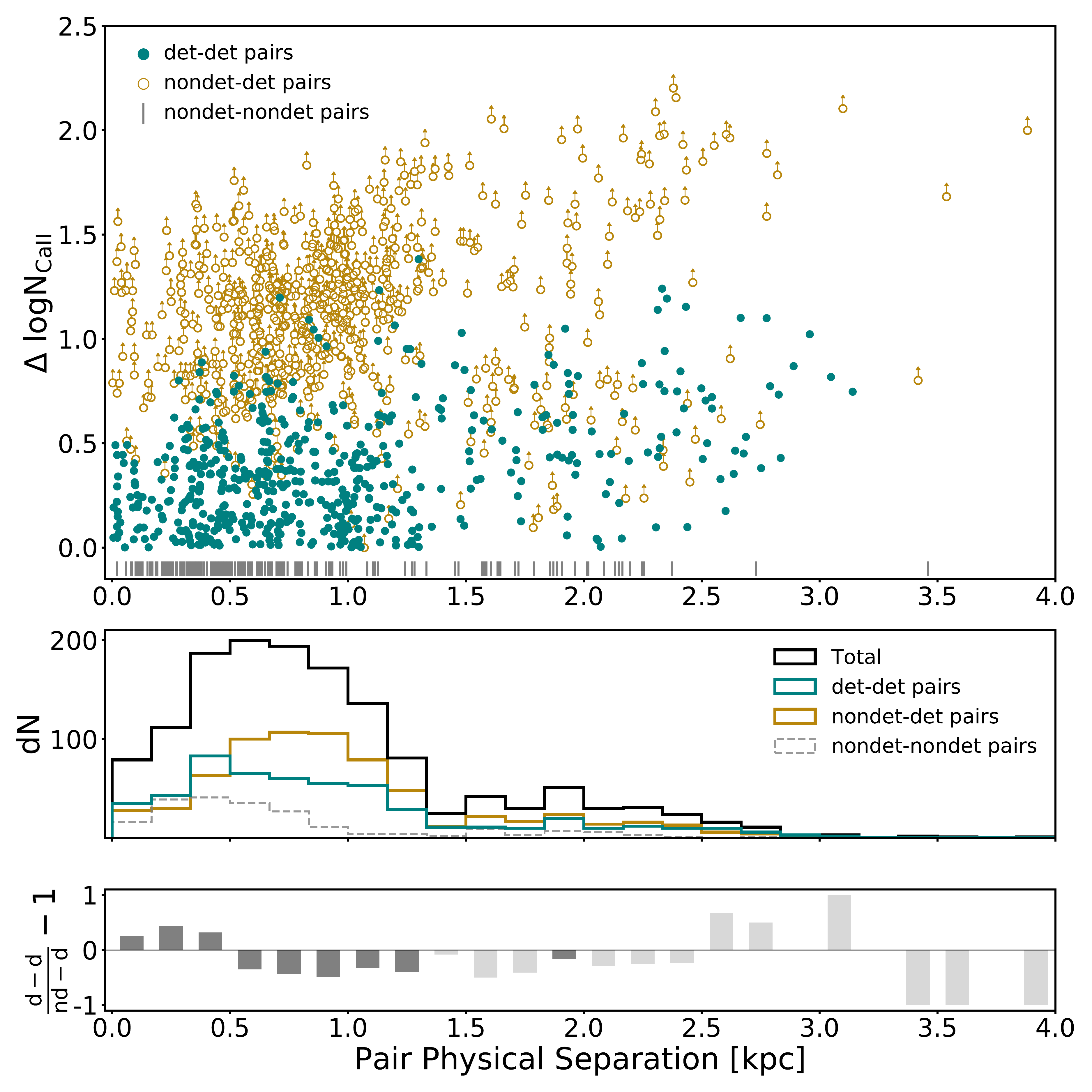}
    \caption{{\sc Column density variations indicate cloud sizes $<$ 500 pc.} The spatial variation of logN, which can place constraints on gas cloud size. The physical separation between IV gas along each unique pair of sightlines is calculated assuming all gas lies at z=3 kpc. \\
    {\it Top:} Filled teal markers represent the difference in CaII column density of gas between two sightlines with CaII detections (d-d). If one sightline in the pair is a detection and one a nondetection (nd-d), lower limits on $\rm \Delta logN$ are determined by instrumental detection limits and indicated with an open gold marker. If both sightlines in a pair are nondetections (nd-nd), their physical separation is marked with a grey bar at the bottom of the figure. \\
    {\it Middle:} The distribution of sightline pair separations is shown in the histogram at the top of the figure. The full set of all sightline pairs is in black, d-d pairs in teal, nd-d pairs in gold, and nd-nd in dashed grey. \\
    {\it Bottom:} The ratio of the number of d-d pairs to nd-d pairs for each bin of the histogram above. The ratio is normalized so that it is positive when there are more d-d pairs and negative when there are more n-d pairs. Light grey bars are shown for bins in which there are fewer than 50 sightline pairs. Below scales of 0.5 kpc, there are more d-d pairs than nd-d pairs, and there is an inversion of this ratio at scales larger than 0.5 kpc. This may be an indication that there is gas substructure at scales on the order of 0.5 kpc.}
    \label{fig:rubinfig}
\end{figure*}

Statistical variations in column density can also provide constraints on the size of gas clouds and the uniformity of their coverage across the disk \citep{rubin2018a,rubin2018b}. To show how column densities vary on different scales, we plot the difference in IV gas column density, $\rm \Delta logN_{CaII}$, between each unique pair of sightlines in Figure \ref{fig:rubinfig}. NaI and HI are not shown, but the results for those species are similar to CaII. We assume that the gas lies at a height of 3 kpc, as suggested by the trends discussed above. Figure \ref{fig:xycloudcoords} provides an example of how the distance scales in question are affected for a range of assumed cloud heights.

In the top panel, each marker represents the difference in column density for two sightlines separated by some physical distance. Pairs of sightlines with two detections (d-d) are represented with filled teal markers. Pairs with one detection and one nondetection (nd-d) are represented by open gold markers showing the lower limits on $\rm \Delta logN$ as determined by instrumental detection limits. The grey bars at the bottom of the plot show the physical separation of sightline pairs with two nondetections (nd-nd). If the spatial sampling density of our sightlines is high enough to probe cloud substructure, we would expect $\rm \Delta logN$ to be smaller at pair separations less than the characteristic clump size. We find that column densities differ by up to two orders of magnitude, but there is no strong evidence of density fluctuations on any particular physical scale.

The histogram in the middle panel displays the distribution of sightline pair separations. The full set of all sightline pairs is in black, d-d pairs in teal, nd-d pairs in gold, and nd-nd in dashed grey. Note that the distribution of d-d pairs is skewed relative to the total distribution of sightline pairs. We expect the number of d-d pairs to increase on scales smaller than typical gas structure.
The bar graph in bottom panel represents the ratio of the number of d-d pairs to nd-d pairs corresponding to each bin of the histogram above. The ratio is normalized so that it is positive when there are more d-d pairs and negative when there are more n-d pairs. At scales shorter than 0.5 kpc, there are more d-d pairs than nd-d pairs, and there is an inversion of this ratio at scales larger than 0.5 kpc.

The absence of a trend at separations larger than 0.5 kpc is an indication that any existing substructure occurs at scales no larger than 0.5 kpc, approximately twice the average separation between a sightline and its nearest neighbor (0.24 kpc). Similarly, large-scale cloud structures are likely to span areas larger than the $\sim 12.6$ kpc$^{2}$ covered by this survey. These constraints are consistent with the HI gas structure shown in Figure \ref{fig:WCS_N}, where dense clumps have formed on scales smaller than our typical sightline separation and make up a cloud complex which extends beyond the survey area.

\subsubsection{Covering Fraction \& Volume Filling Factor}
\label{sec:fillingfactor}
We calculate covering fraction as a convenient and simple measure of the projected sky coverage and distribution of IV gas. Before proceeding with the calculation we must recognize that our sightlines extend to varying distances, and this should be taken into account. In this case, we assume that the detected gas lies at $z=3$ kpc and therefore all sightlines fully probe these gas clouds, and assert that this is a valid assumption based on the argument laid out in \S\ref{sec:columndensity}. See Figure \ref{fig:xycloudcoords} for an illustration of the effect of cloud height assumptions on distance scales.

We must also choose a column density threshold above which the covering fraction will apply. Column density measurements are a mix of detections and upper limits representing non-detections. Ideally there should be no upper limits above the chosen threshold, since this does not allow us to categorize that sightline definitively as a detection or non-detection. However, choosing a threshold that meets this criterion would exclude a significant fraction of detections in our sample (see Figure \ref{fig:NvsZ}). In light of this we take the following approach: we choose a threshold of logN$_{\rm IV}$ $>$ 11.5 for CaII, which includes all detections and exceeds all but four upper limits, and a threshold of logN$_{\rm IV}$ $>$ 11.3 for NaI, which includes all detections and exceeds all but one upper limit. We treat the upper limits which exceed the threshold as detections and account for this uncertainty in the covering fraction error. The covering fraction error is determined by a standard binomial Wilson score; we calculate the 68\% confidence interval treating the upper limits above the threshold as both detections and non-detections, and report the most conservative of those two values.

Using 54 sightlines, we calculate covering fractions of $f_{\rm CaII}$ ($\rm log N_{CaII}$ $>$ 11.5) $= 63^{+6.5}_{-14.0}\%$, ~ $f_{\rm NaI}$ ($\rm log N_{NaI}$ $>$ 11.3) $= 26^{+5.9}_{-7.6}\%$, and $f_{\rm HI}$ ($\rm log N_{HI}$ $>$ 18.7) $= 52\pm 5.8\%$.
Other studies of neutral HI clouds in the Milky Way have found smaller covering fractions of $\sim$20-40\% \citep{murphy1995,wakker1991,albertdanly2004}. Studies of warm HVCs in the Milky Way halo find covering fractions of $\sim$60-90\%, larger than the low-ionization covering fractions we measure \citep{shull2009,lehnerhowk2011,collins2009}. 

If we use our statistical analysis in \S\ref{sec:cloudsizeanddist} to make an assumption about cloud size, we can also calculate the volume filling factor, given by

\begin{equation} \label{eq1}
\begin{split}
F & = N_{c}\frac{\frac{4}{3}\pi r_c^3}{\pi R_s^2z} \\
\end{split}
\end{equation}

$N_{c}$ is the total number of IV components detected along all sightlines, including six sightlines with two IV components and one sightline with three IV components. We assume that each IV gas detection is a distinct spherical cloud substructure, and that the size of these structures must be smaller than the scales we probe with this survey given the absence of statistical variation in column density described in \S\ref{sec:cloudsizeanddist}. We assign an upper limit of $r_c \leq 0.1225$ kpc on cloud radius using the average $xy$-projected distance between a sightline and its nearest neighbor ($0.245$ kpc). $\pi R_s^2=12.6$ kpc$^2$ is the disk area covered by the cylindrical survey volume, and we assume the height of that cylinder to be $z=3$ kpc since we find no evidence of IV gas at $z>3$ kpc. This gives us volume filling factors of $F_{\rm CaII}$ ($\rm log N_{CaII}$ $>$ 11.5) $\leq 0.86^{+0.07}_{-0.15}\%$, $F_{\rm NaI}$ ($\rm log N_{NaI}$ $>$ 11.3) $\leq 0.35^{+0.06}_{-0.08}\%$, and $F_{\rm HI}$ ($\rm log N_{HI}$ $>$ 18.7) $\leq 0.63\pm 0.06\%$.

Although these volume filling factors are quite small compared to the covering fraction we calculate ($F_{\rm CaII} \leq 0.28\%$ vs. $f_{\rm CaII} = 63\%$), the values are not contradictory if cloud sizes are small, as suggested by our analysis of column density fluctuations in \S\ref{sec:cloudsizeanddist}. Even for length scales as short as $\sim 1$ kpc the covering fraction of small clouds easily reaches unity, which can explain the high detection rates of low-ion CGM gas if it does in fact clump on scales $< 500$ pc \citep{liang2018}.
\\

\section{Discussion}
\label{sec:discussion}

\subsection{Implications of Small Cloud Size}
\label{sec:implications}
The larger covering fractions and volume filling factors we see for higher ioinization states support a multi-phase picture in which small clumps or droplets of cool gas are suspended within larger clouds or streams of hotter, more diffuse gas. How this gas eventually cools and becomes incorporated into the interstellar medium remains controversial. At $T \sim 10^6$ K, thermal line emission can rapidly cool gas until it reaches $T \sim 10^4$ K, where the cooling curve drops off steeply \citep{sutherlanddopita1993,maio2007}. \cite{mallerbullock2004} argue that thermal instabilities naturally arise in the Galactic corona in which cool clouds condense from a hot medium, while \cite{binney2009} have pointed out that the halo, under most assumed physical conditions, is largely stable to cooling. Instead, it may be that feedback-induced or fountain-induced cooling, in which metal-enriched clouds ejected from the disk mix with coronal material and enhance the subsequent cooling, are the relevant processes driving accretion at the disk-halo interface \citep{marinacci2010,marinacci2012, fraternali2017, howk2018}. Detailed, global hydrodynamic disk simulations will be critical to resolving the physics of gas accretion and the Galactic fountain \citep[e.g.][]{fielding2017, schneider2018, kimostriker2018}.

The largely empirical picture of cooler, denser gas embedded within warmer, more diffuse clouds may provide insight as to why we have more sparse detections in NaI, which predominantly traces dust and dense-neutral gas ($T \lesssim$ 1,000 K), than CaII, which traces both dense-neutral and warmer ionized gas ($T \lesssim$ 10,000 K), despite the fact that our sensitivity to both is roughly the same \citep{murga2015,puspitarinilallement2012}. If NaI clouds sit within larger CaII clouds in such a hierarchical structure, then some random sightline is less likely to pass through the NaI cloud \citep{stern2016}. More generally, if the cold phase of CGM gas lies in small, clumpy structures rather than diffuse clouds that fill a larger volume in warm phase gas, then low-ionization clouds must be relatively common in order to explain the detection rates we see.

Currently, most galaxy simulations can resolve scales of $\sim$500 pc at best and usually do not focus computational resources on resolving low-density gas in the CGM. In general these simulations underpredict the amount of observed CGM ions \citep[e.g.][]{peeples2018, hummels2018}. Incorporating AGN feedback has improved predictions of high-ion abundances, but simulations still struggle to reproduce low-ion abundances, even for hydrogen \citep{oppenheimer2018}. If low-ion CGM clouds do form on scales $< 500$ pc as our analysis and that of others suggests, then the resolution of CGM gas in simulations must increase by orders of magnitude in order to reflect the hydrodynamics on relevant scales \citep{marinacci2010,crighton2015,mccourt2018,liang2018,sparre2019}. Recent work has demonstrated that improving resolution to $< 500$ pc in simulations dramatically changes the characteristics of cool gas clouds \citep{hummels2018,peeples2018}. As simulations improve and move closer to resolving these structures on stabilization scales, observational constraints on cloud properties like the ones we present here will be useful for testing the accuracy of their predictions.
\\

\subsection{Gas Origins \& The Galactic Fountain}
\label{sec:gasorigins}
\label{sec:galacticfountain}

The galactic fountain broadly depicts a scenario in which hot gas is ejected from a galaxy by energetic feedback processes associated with star formation and/or a central supermassive black hole, then re-accretes onto the disk to catalyze new star formation in an ongoing cycle.
Our findings indicate a substantial amount of cool extraplanar gas embedded within neutral structures moving toward the disk at distances less than 3.1 kpc. The observed bulk kinematics presented in this study provide clues about the origin and fate of gas at the disk-halo interface. In this section we consider our observations in the context of the `galactic fountain' and the potential origins of neutral infalling gas. In particular, we examine how the data might be explained by gas ejected from elsewhere within the Milky Way: either by star formation processes throughout the disk, or driven by a central galactic engine. A viable model should be able to reproduce the velocity gradient and lack of column density trends we observe.

One well-established picture of the galactic fountain places the origin of infalling gas at the sites of supernovae explosions throughout the disk. \cite{fraternali2017} models several variations of the disk-ejection fountain scenario that could produce the velocity gradient we see: in a simple ballistic model, particles ejected vertically from the disk will fall back to the disk at larger radii because of conservation of angular momentum. Since gravitational potential decreases with galactic radius, ejecta at larger radii will reach a greater height. In turn, this material will reach more negative velocities by the time it reaches the disk. Such an effect could produce a radius-dependent infall velocity gradient for cooled, infalling gas. Adding in effects of condensation and drag could mean that these particles also return to the disk at significantly smaller radii. Our data can be explained by a disk-ejection scenario if outflowing gas is in a hotter phase that is difficult to detect with low ion absorption, or if the outflows are time-variable over many Myr as a result of intermittent, clustered star formation, as in the models of \cite{kimostriker2018}. Although cool outflowing gas has been detected in other local galaxies \citep{heckman2000,martin2005,rupke2005,chen2010,rubin2014,bordoloi2014a}, such outflows may have a patchy distribution across the disk, over which our observations span only a small fraction.

The velocity gradient we observe across the disk could potentially emerge in a central engine scenario as well \citep{bordoloi2017,fox2015a}. In this picture, gas is driven out from the center of the galaxy in a roughly conical region extending above the disk. Gas clouds cooling and condensing at the boundary of the cone at higher galactic radii would also be at greater height above the disk. This would allow gas clouds at greater radii to reach greater velocities by the time they reach the disk-halo interface. The effect would be a similar radius-dependent velocity gradient to the one described for the disk-ejection scenario above, but through a different mechanism. Over time the trajectory of this infalling gas turns toward the potential well at galactic center and begins to move more quickly across the plane of the disk, which is consistent with a lack of detected outflows in the solar neighborhood. Gas clouds which condense further from galactic center would have longer infall times and therefore experience a greater trajectory change, leading to a decrease in $z$-component velocities as they fall and turn towards galactic center. Although this observed phenomenon emerges nicely in a simple toy model of this scenario, it does not easily reproduce gas that predominantly lies at small $z$ unless a full opening angle of $\phi > 110^{\circ}$ is adopted for initial gas ejection. A larger sample of halo star sightlines over the full sky would be necessary to fully test this proposed scenario.

In principle, the metal abundance of the clouds we observe could also place constraints on its origin. Gas with low metal abundance may have originated from the IGM, or been stripped from the ISM of Milky Way satellite galaxies \citep{kirby2011}. Larger metal abundances would suggest that material has been processed through stars in the disk and ejected via outflows as part of a galactic fountain. Furthermore, the differing elemental yields of sources like supernovae, stellar winds, or AGN could potentially be tied to the relative abundances of various species across time scales of $\sim$100-300 Myr \citep{krumholz2018,emerick2018}. Unfortunately, determining the metal abundance for the extraplanar gas we observe is overall a very problematic calculation, and our data do not independently constrain the gas metallicity because of the lack of HI in absorption. We refer the reader to \citet{wakker2001} and \citet{richter2001b} for discussions of metallicity determination in the IV Arch, which is estimated to be approximately solar.

When examining these results in the context of the galactic fountain, it is important to note that our velocity measurements reflect only local conditions and may not be representative of fountain gas in a global sense. Although the vertical gas velocities we measure are constraining in their own right, our sightlines do not capture gas flows in the transverse direction which are also critical for understanding fountain kinematics. However, our well-sampled kinematic measurements reinforce and confirm the observed properties of intermediate-velocity HI clouds, which have been successfully reproduced by galactic fountain models \citep{marasco2011,marasco2012}. The dynamics of halo gas in the Milky Way are complex, and these scenarios are highly simplified pictures which we have discussed in isolation. In reality gas kinematics are driven by an interrelated combination of several mechanisms which continuously supply the disk with gas.
\newline
\newline

\section{Summary}
\label{sec:summary}
 
In this work we have performed an analysis of cool ($T \sim 10^4$ K) gas at the Milky Way's disk-halo interface by examining absorption features of intermediate-velocity ($25 < |v| < 75$ km/s) CaII and NaI along sightlines to 54 high-latitude blue horizontal branch (BHB) stars at heights of 3.1-13.4 kpc above the disk in the northern Galactic hemisphere. Distance measurements and dense sightline sampling allow us to obtain a 3-dimensional picture of the spatial extent of the gas, which we combine with the excellent resolution of Keck HIRES spectra and compare to HI 21-cm emission maps to constrain the characteristics of CGM gas flows.

\begin{enumerate}
    \item CaII, and to a lesser extent the more sparsely detected NaI, are spatially coincident with HI emission. The detections are grouped together in the direction away from Galactic center at the edge of two large HI complexes known as the IV Arch and the IV Spur. However, CaII and HI column densities are not significantly correlated, which may be an observational effect of the large beam size of the EBHIS HI survey, or an indication that HI is associated with gas in a more highly ionized phase. (\S\ref{sec:columndensity} and \S\ref{sec:discussion}; Figures \ref{fig:IVcol} \& \ref{fig:WCS_N})
    
    \item We find no relationship between column density along a sightline and distance to the background source, indicating that these gas clouds lie close to the disk at $z \lesssim 3.1$ kpc. (\S\ref{sec:columndensity}; Figure \ref{fig:NvsZ})
    
    \item We detect virtually no outflowing gas, and inflows at velocities of $-75$ $< v_z <$ $-25$ km/s. The gas exhibits a clear velocity gradient of $\sim$6-9 km/s/kpc across the disk which does not correlate with the column density of the gas. (\S\ref{sec:kinematics}; Figures \ref{fig:IVvel} \& \ref{fig:veltrendfits})
    
    \item An analysis of statistical variations in column density suggests that substructure of low ions within warm clouds exists on scales $<$ 500 pc. Most current hydrodynamical simulations do not resolve cool CGM gas on these scales, and greater resolution may be crucial for improving simulations' chronically low predictions for observed low-ion column densities. (\S\ref{sec:cloudsizeanddist}; Figure \ref{fig:rubinfig})
    
    \item We calculate covering fractions of $f_{\rm CaII} = 63^{+6.5}_{-14.0}\%$, ~ $f_{\rm NaI} = 26^{+5.9}_{-7.6}\%$, and $f_{\rm HI} = 52\pm 5.8\%$, and volume filling factors of $F_{\rm CaII} \leq 0.86^{+0.07}_{-0.15}\%$, $F_{\rm NaI} \leq 0.35^{+0.06}_{-0.08}\%$, and $F_{\rm HI} \leq 0.63\pm 0.06\%$. A comparison of covering fractions for different ions supports a multi-phase picture in which small clumps or droplets of cool gas are suspended within larger clouds or streams of hotter, more diffuse gas. (\S\ref{sec:fillingfactor})
    \\
    
\end{enumerate}

\acknowledgements
   
Support for this work was provided by NASA through program GO-14140. JKW acknowledges support from a 2018 Sloan Foundation Fellowship, and from the Research Royalty Fund Grant 65-5743 at the University of Washington. AD is supported by a Royal Society University Research Fellowship. 
The optical data presented herein were obtained at the W. M. Keck Observatory, which is operated as a scientific partnership among the California Institute of Technology, the University of California and the National Aeronautics and Space Administration. The Observatory was made possible by the generous financial support of the W. M. Keck Foundation. The authors wish to recognize and acknowledge the very significant cultural role and reverence that the summit of Maunakea has always had within the indigenous Hawaiian community.  We are most fortunate to have the opportunity to conduct observations from this mountain. 
We thank Chris Howk, Mary Putman, Josh Peek, Karin Sandstrom, Cameron Hummels, Joe Hennawi, and Drummond Fielding for helpful input and discussions about the project. HVB thanks the anonymous referee for numerous suggestions that improved this work. We would also like to acknowledge the participants of the 2018 Arthur M. Wolfe Symposium in Astrophysics and their role in the exchange of ideas that made this work possible.  Finally, AD and JKW thank Risa Wechsler and Charlie Conroy for organizing a speed collaboration session at the Mayacamas Ranch Conference in 2015, where the idea for this project originally took shape.

\facility{Keck: HIRES}


\bibliographystyle{apj}
\bibliography{apjmanuscript_BHBs.bbl}
\clearpage

\begin{deluxetable*}{l c c c c r c c c r}

\tabletypesize{\scriptsize}
\tablecaption{BHB Background Sources}
\tablehead{
\colhead{Name} &
\colhead{RA} &
\colhead{Dec} &
\colhead{$x$} &
\colhead{$y$} &
\colhead{$z$} &
\colhead{Distance} &
\colhead{$b$} &
\colhead{$l$} &
\colhead{HRV}\\
& 
\colhead{[h:m:s]} & 
\colhead{[$^{\circ}$:$\arcmin$:$\arcsec$]} & 
\colhead{[kpc]} & 
\colhead{[kpc]} & 
\colhead{[kpc]} & 
\colhead{[kpc]} & 
\colhead{[$^{\circ}$]} & 
\colhead{[$^{\circ}$]} & 
\colhead{[km/s]}
}
\startdata
J1149+2828 & 11:49:19.16  & +28:28:06.7 & 9.8 & \ 0.8  & \ 8.2  &  8.5  & 76.2  & 203.8 & 102.9\\
J1204+1947 & 12:04:05.42  & +19:47:43.6 & 8.7 & \ 1.5  & \ 7.1  &  7.3  & 76.9  & 244.5 & -205.2\\
J1212+2826 & 12:12:18.09  & +28:26:02.0 & 9.7 & \ 0.7  &  12.1  & 12.2  & 81.3  & 202.1 & 40.6\\
J1215+2315 & 12:15:59.62  & +23:15:57.9 & 8.8 & \ 1.2  & \ 9.5  &  9.6  & 81.1  & 236.7 & 61.9\\
J1215+3202 & 12:15:01.55  & +32:02:20.1 & 9.6 & \ 0.0  & \ 9.5  &  9.7  & 80.7  & 178.9 & -88.4\\
J1216+3308$^{\dag}$ & 12:16:00.06  & +33:08:56.2 & 9.4 &   -0.2  &   8.4  &  8.6 & 80.3 & 172.6 & 39.8\\
J1216+3004 & 12:16:53.67  & +30:04:08.7 & 9.6 & \ 0.3  &  11.5  & 11.6  & 81.9  & 189.7 & 59.2\\
J1217+2104 & 12:17:21.04  & +21:04:11.1 & 8.6 & \ 1.6  & \ 9.7  &  9.9  & 80.2  & 249.0 & -129.4\\
J1218+2951 & 12:18:00.44  & +29:51:38.3 & 9.4 & \ 0.3  &  10.6  & 10.7  & 82.2  & 190.6 & -166.9\\
J1219+2559 & 12:19:25.79  & +25:59:43.0 & 8.7 & \ 0.6  & \ 6.9  &  7.0  & 82.8  & 220.1 & 174.2\\
J1223+0002 & 12:23:55.21  & +00:02:20.9 & 7.4 & \ 2.0  & \ 3.9  &  4.4  & 62.1  & 288.1 & 139.8\\
J1223+3641 & 12:23:26.85  & +36:41:11.1 & 9.5 &  -0.8  & \ 8.6  &  8.7  & 78.8  & 153.0 & -80.4\\
J1226+1749 & 12:26:45.86  & +17:49:52.0 & 8.0 & \ 1.9  & \ 9.6  &  9.8  & 79.1  & 270.2 & 60.5\\
J1229+2101 & 12:29:47.49  & +21:01:35.9 & 8.2 & \ 1.9  &  14.1  & 14.2  & 82.2  & 262.8 & 211.2\\
J1230+2052 & 12:30:22.97  & +20:52:21.9 & 8.2 & \ 1.8  &  13.4  & 13.5  & 82.1  & 264.3 & 174.2\\
J1231+3719 & 12:31:10.44  & +37:19:08.4 & 9.4 &  -1.0  & \ 8.8  &  9.0  & 79.0  & 144.4 & -168.0\\
J1234+4437 & 12:34:32.86  & +44:37:32.1 & 9.4 &  -1.5  & \ 6.4  &  6.7  & 72.2  & 132.8 & 48.7\\
J1236+2837 & 12:36:51.75  & +28:37:36.0 & 8.6 & \ 0.1  &  10.1  & 10.1  & 86.5  & 187.1 & -195.8\\
J1239+2937 & 12:39:18.13  & +29:37:52.8 & 8.3 &  -0.1  & \ 4.1  &  4.1  & 86.3  & 169.1 & 58.8\\
J1240+2746 & 12:40:54.84  & +27:46:40.6 & 8.3 & \ 0.1  & \ 7.7  &  7.7  & 87.6  & 196.8 & 111.6\\
J1241+2909 & 12:41:17.48  & +29:09:14.4 & 8.2 & \ 0.0  & \ 4.3  &  4.3  & 87.0  & 170.2 & -69.9\\
J1242+2126 & 12:42:03.34  & +21:26:04.2 & 7.8 & \ 0.8  & \ 7.8  &  7.8  & 83.9  & 281.9 & -71.3\\
J1243+2842 & 12:43:50.94  & +28:42:35.8 & 8.3 &  -0.1  & \ 7.6  &  7.6  & 87.7  & 169.2 & -102.8\\
J1243+2931 & 12:43:39.02  & +29:31:49.1 & 8.2 &  -0.1  & \ 4.9  &  4.9  & 87.0  & 158.0 & 17.2\\
J1244+2926 & 12:44:18.54  & +29:26:19.4 & 8.5 &  -0.2  &  11.2  & 11.2  & 87.2  & 156.7 & 99.5\\
J1244+2939 & 12:44:16.47  & +29:39:45.3 & 8.5 &  -0.2  &  10.5  & 10.5  & 87.0  & 154.4 & -191.6\\
J1248+0947 & 12:48:10.28  & +09:47:34.4 & 7.2 & \ 1.3  & \ 4.9  &  5.1  & 72.7  & 300.2 & 80.2\\
J1251+3849 & 12:51:11.16  & +38:49:54.1 & 8.8 &  -1.3  & \ 7.3  &  7.4  & 78.3  & 123.2 & -104.7\\
J1255+0924 & 12:55:42.55  & +09:24:23.4 & 7.0 & \ 1.3  & \ 5.2  &  5.5  & 72.3  & 306.4 & 150.2\\
J1257+1741$^{\dag}$ & 12:57:57.64  & +17:41:51.1 & 7.0 & \ 1.1 & 9.2 & 9.3 & 80.4 & 312.3 & 41.3\\
J1258+1939 & 12:58:58.86  & +19:39:29.1 & 6.7 & \ 1.2  &  13.0  & 13.1  & 82.3  & 316.4 & -162.3\\
J1259+1552 & 12:59:53.71  & +15:52:27.7 & 6.9 & \ 1.2  & \ 8.3  &  8.5  & 78.6  & 313.3 & 47.1\\
J1306+2003 & 13:06:25.42  & +20:03:42.3 & 6.5 & \ 0.9  &  12.8  & 12.9  & 82.1  & 329.6 & 27.4\\
J1307+2224 & 13:07:00.15  & +22:24:25.7 & 6.7 & \ 0.4  &  13.0  & 13.1  & 84.1  & 340.6 & 187.9\\
J1310+2514 & 13:10:40.39  & +25:14:17.4 & 7.2 &  -0.1  & \ 9.6  &  9.6  & 85.2  &  10.4 & 38.4\\
J1314+1751 & 13:14:13.24  & +17:51:57.6 & 6.4 & \ 0.8  & \ 9.3  &  9.5  & 79.4  & 333.7 & -31.3\\
J1324+2038 & 13:24:36.40  & +20:38:17.1 & 6.5 & \ 0.2  & \ 8.3  &  8.4  & 80.0  & 354.1 & 28.5\\
J1324+2418 & 13:24:53.36  & +24:18:36.2 & 6.4 &  -0.4  &  12.0  & 12.1  & 82.0  &  14.3 & 100.1\\
J1325+2232a & 13:25:26.16 & +22:32:20.5 & 6.5 &  -0.1  & \ 9.5  &  9.6  & 81.0  &   4.0 & -148.4\\
J1325+2232b & 13:25:54.40 & +22:32:50.1 & 6.8 &  -0.1  & \ 7.3  &  7.4  & 80.9  &   4.4 & -100.3\\
J1332+2054 & 13:32:56.98  & +20:54:41.5 & 6.1 &  -0.1  & \ 9.5  &  9.7  & 78.7  &   1.9 & 109.7\\
J1335+2820 & 13:35:47.01  & +28:20:41.2 & 6.7 &  -1.2  &  10.4  & 10.6  & 80.1  &  42.6 & 135.3\\
J1338+2345 & 13:38:25.33  & +23:45:51.6 & 6.2 &  -0.6  & \ 9.9  & 10.1  & 78.9  &  17.9 & -87.8\\
J1341+2801 & 13:41:18.85  & +28:01:56.3 & 6.5 &  -1.3  & \ 9.9  & 10.1  & 78.9  &  40.5 & -142.4\\
J1341+2806 & 13:41:52.29  & +28:06:04.3 & 6.9 &  -0.9  & \ 7.3  &  7.5  & 78.8  &  40.8 & -166.8\\
J1341+2823 & 13:41:19.85  & +28:23:58.6 & 6.4 &  -1.5  &  11.1  & 11.3  & 78.9  &  42.4 & -124.8\\
J1341+2824 & 13:41:27.90  & +28:24:29.9 & 6.5 &  -1.3  &  10.0  & 10.2  & 78.9  &  42.4 & -133.4\\
J1341+2829 & 13:41:38.30  & +28:29:54.4 & 6.6 &  -1.3  &  10.0  & 10.2  & 78.8  &  42.9 & -134.6\\
J1342+2828 & 13:42:43.32  & +28:28:15.8 & 6.6 &  -1.3  & \ 9.6  &  9.8  & 78.6  &  42.7 & -128.8\\
J1344+1842 & 13:44:04.41  & +18:42:59.1 & 6.5 & \ 0.0  & \ 5.7  &  5.9  & 75.3  &   0.9 & -74.7\\
J1347+1811 & 13:47:49.70  & +18:11:43.6 & 6.5 & \ 0.0  & \ 5.3  &  5.5  & 74.2  &   1.4 & 18.2\\
J1413+5621 & 14:13:37.73  & +56:21:57.7 & 8.5 &  -2.2  & \ 3.5  &  4.1  & 57.3  & 101.9 & 20.2\\
J1415+3716 & 14:15:57.13  & +37:16:58.1 & 6.7 &  -3.2  & \ 9.2  &  9.9  & 69.5  &  67.9 & -134.0\\
J1420+5520 & 14:20:43.45  & +55:20:35.0 & 8.3 &  -1.9  & \ 3.1  &  3.7  & 57.5  &  99.2 & -89.3\\
J1527+4027 & 15:27:50.05  & +40:27:28.9 & 7.0 &  -2.1  & \ 3.4  &  4.1  & 55.2  &  65.6 & -105.6\\
J1534+5015 & 15:34:11.23  & +50:15:56.3 & 7.5 &  -3.2  & \ 4.1  &  5.2  & 51.4  &  81.0 & -129.4\\
\enddata
\tablecomments{Properties of blue horizontal branch stars (BHBs) used as background sources to measure gas absorption. The BHB coordinates are given for a Cartesian system in which the $x$-axis extends from Galactic center in the direction of the Sun, $y$ is perpendicular to $x$ in the plane of the disk, and $z$ is perpendicular to the disk such that the system is right-handed, with positive $z$ towards the north Galactic pole (see \S\ref{sec:coordinatesystem}). $HRV$ is the heliocentric radial velocity of the BHB and $b$ and $l$ are Galactic latitude and longitude, respectively. Coordinates and distance measurements were obtained from the SDSS SEGUE catalog \citep{xue2011}. Errors on distance measurements are approximately 10\%. \\
$^{\dag}$Excluded from sample because of poor data quality \\}
\label{table:bhblist}
\end{deluxetable*}


\clearpage

\LongTables

\begin{deluxetable*}{l || c c c | c c r}
\tabletypesize{\scriptsize}
\tablecaption{MCMC Best-fit Absorption Line Properties}
\tablehead{
 &
 &
\colhead{\bf CaII} &
 &
 &
\colhead{\bf NaI} &
 \\
\colhead{Name} & 
\colhead{$\rm v_{LSR}$ [km/s]} & 
\colhead{$\rm logN$} & 
\colhead{b [km/s]} & 
\colhead{$\rm v_{LSR}$ [km/s]} & 
\colhead{$\rm logN$} & 
\colhead{b [km/s]} 
}
\startdata
J1149+2828a & $-47.5\,_{-0.6}^{+0.6}$ & $12.22\,_{-0.03}^{+0.03}$ & $ 9.5\,_{- 1.1}^{+ 1.1}$    & $-49.0\,_{-2.7}^{+3.5}$ & $11.38\,_{-0.13}^{+0.20}$ & $13.5\,_{- 4.1}^{+14.9}$  \\ [1ex] 
            & $-20.9\,_{-0.4}^{+0.4}$ & $12.07\,_{-0.23}^{+0.45}$ & $ 1.2\,_{- 0.4}^{+ 1.4}$    & $-19.7\,_{-0.6}^{+0.7}$ & $11.30\,_{-0.08}^{+0.08}$ & $ 2.0\,_{- 1.1}^{+ 1.5}$  \\ [3ex] 
\\  
J1204+1947a & $-40.6\,_{-1.1}^{+1.5}$ & $12.03\,_{-0.06}^{+0.07}$ & $12.2\,_{- 2.6}^{+ 3.9}$    & $ -8.1\,_{-2.4}^{+1.5}$ & $11.38\,_{-0.19}^{+0.13}$ & $ 4.7\,_{- 2.5}^{+ 3.7}$  \\ [1ex] 
            & $-12.7\,_{-0.6}^{+0.6}$ & $12.25\,_{-0.03}^{+0.03}$ & $ 9.9\,_{- 1.1}^{+ 1.1}$    &    --                   &     --                    &    --                     \\ [3ex] 
\\  
J1212+2826a & $-47.6\,_{-1.8}^{+2.0}$ & $12.40\,_{-0.05}^{+0.05}$ & $19.9\,_{- 2.3}^{+ 2.9}$    &    --                   &     --                    &    --                     \\ [1ex] 
            & $-19.6\,_{-0.5}^{+0.5}$ & $11.83\,_{-0.04}^{+0.04}$ & $ 5.3\,_{- 1.0}^{+ 1.0}$    &    --                   &     --                    &    --                     \\ [3ex] 
\\  
J1215+2315a & $-43.1\,_{-1.3}^{+1.5}$ & $12.18\,_{-0.05}^{+0.06}$ & $16.5\,_{- 2.4}^{+ 3.6}$    & $-47.2\,_{-2.4}^{+3.8}$ & $11.40\,_{-0.13}^{+0.30}$ & $14.3\,_{- 4.7}^{+22.7}$  \\ [1ex] 
            & $-20.1\,_{-0.6}^{+0.5}$ & $12.15\,_{-0.03}^{+0.04}$ & $ 8.0\,_{- 0.9}^{+ 1.0}$    & $-20.8\,_{-0.1}^{+0.1}$ & $12.23\,_{-0.04}^{+0.06}$ & $ 2.7\,_{- 0.3}^{+ 0.3}$  \\ [3ex] 
\\  
J1215+3202a & $-47.8\,_{-0.8}^{+0.7}$ & $12.19\,_{-0.03}^{+0.03}$ & $11.3\,_{- 1.2}^{+ 1.6}$    & $-16.0\,_{-0.3}^{+0.4}$ & $11.33\,_{-0.05}^{+0.06}$ & $ 1.6\,_{- 0.5}^{+ 0.7}$  \\ [1ex] 
            & $-16.9\,_{-1.0}^{+1.1}$ & $12.02\,_{-0.05}^{+0.06}$ & $10.9\,_{- 1.8}^{+ 2.5}$    &    --                   &     --                    &    --                     \\ [3ex] 
\\  
J1216+3004a & $-44.7\,_{-0.4}^{+0.4}$ & $12.29\,_{-0.02}^{+0.02}$ & $ 9.8\,_{- 0.8}^{+ 0.8}$    & $-43.6\,_{-0.1}^{+0.2}$ & $12.03\,_{-0.07}^{+0.06}$ & $ 1.9\,_{- 0.2}^{+ 0.4}$  \\ [1ex] 
            & $-10.4\,_{-1.0}^{+1.5}$ & $12.30\,_{-0.05}^{+0.06}$ & $18.6\,_{- 2.4}^{+ 3.5}$    & $-14.0\,_{-0.2}^{+0.2}$ & $11.60\,_{-0.02}^{+0.02}$ & $ 3.8\,_{- 0.5}^{+ 0.5}$  \\ [3ex] 
\\  
J1217+2104a & $-44.9\,_{-1.1}^{+1.0}$ & $12.09\,_{-0.06}^{+0.07}$ & $10.6\,_{- 1.6}^{+ 3.0}$    & $ -0.8\,_{-0.5}^{+0.5}$ & $11.85\,_{-0.07}^{+0.07}$ & $ 4.2\,_{- 0.8}^{+ 0.8}$  \\ [1ex] 
            & $-24.3\,_{-1.0}^{+1.0}$ & $11.97\,_{-0.07}^{+0.10}$ & $ 7.0\,_{- 1.5}^{+ 3.9}$    &    --                   &     --                    &    --                     \\ [1ex] 
            & $  1.2\,_{-0.7}^{+0.7}$ & $12.47\,_{-0.02}^{+0.03}$ & $14.3\,_{- 1.0}^{+ 1.2}$    &    --                   &     --                    &    --                     \\ [3ex] 
\\  
J1218+2951a & $-44.4\,_{-0.5}^{+0.5}$ & $12.24\,_{-0.04}^{+0.03}$ & $ 8.7\,_{- 0.7}^{+ 0.8}$    & $-46.0\,_{-1.5}^{+1.3}$ & $11.53\,_{-0.07}^{+0.07}$ & $ 8.0\,_{- 1.9}^{+ 2.5}$  \\ [1ex] 
            & $ -6.6\,_{-3.8}^{+3.9}$ & $12.39\,_{-0.24}^{+0.28}$ & $23.5\,_{-11.1}^{+23.2}$    & $-17.8\,_{-0.5}^{+0.4}$ & $11.81\,_{-0.04}^{+0.04}$ & $ 4.5\,_{- 1.0}^{+ 1.1}$  \\ [3ex] 
\\  
J1219+2559a & $-39.4\,_{-0.3}^{+0.3}$ & $12.43\,_{-0.02}^{+0.02}$ & $ 9.7\,_{- 0.5}^{+ 0.5}$    & $-44.0\,_{-0.9}^{+1.2}$ & $11.74\,_{-0.06}^{+0.06}$ & $ 9.1\,_{- 1.5}^{+ 2.4}$  \\ [1ex] 
            & $-14.1\,_{-2.1}^{+1.3}$ & $12.15\,_{-0.07}^{+0.10}$ & $17.6\,_{- 3.5}^{+ 6.7}$    & $-34.0\,_{-1.0}^{+0.9}$ & $11.72\,_{-0.06}^{+0.11}$ & $ 7.8\,_{- 1.7}^{+ 3.9}$  \\ [1ex] 
            &    --                   &     --                    &   --                        & $ -5.3\,_{-1.1}^{+1.0}$ & $11.63\,_{-0.07}^{+0.05}$ & $ 7.8\,_{- 1.7}^{+ 2.2}$  \\ [3ex]   
\\  
J1223+0002a & $-15.1\,_{-1.6}^{+1.5}$ & $12.12\,_{-0.05}^{+0.05}$ & $21.3\,_{- 2.6}^{+ 4.2}$    & $ -7.0\,_{-0.2}^{+0.2}$ & $11.85\,_{-0.09}^{+0.15}$ & $ 1.3\,_{- 0.3}^{+ 0.3}$  \\ [1ex] 
            & $  26.1\,_{-2.3}^{+4.3}$ & $12.10\,_{-0.10}^{+0.19}$ & $23.4\,_{- 6.3}^{+15.5}$    &    --                   &     --                    &    --                     \\ [3ex] 
\\  
J1223+3641a & $-52.7\,_{-1.7}^{+1.0}$ & $12.07\,_{-0.07}^{+0.11}$ & $ 8.3\,_{- 2.1}^{+ 3.1}$    & $-50.6\,_{-0.7}^{+0.7}$ & $11.19\,_{-0.09}^{+0.13}$ & $ 1.7\,_{- 1.1}^{+ 3.9}$  \\ [1ex] 
            & $-16.4\,_{-3.6}^{+1.7}$ & $11.97\,_{-0.11}^{+0.19}$ & $12.7\,_{- 4.3}^{+ 9.1}$    & $-26.5\,_{-0.3}^{+0.3}$ & $11.52\,_{-0.04}^{+0.09}$ & $ 2.1\,_{- 0.9}^{+ 1.2}$  \\ [1ex] 
            & $ 16.8\,_{-0.8}^{+0.8}$ & $11.63\,_{-0.07}^{+0.09}$ & $ 3.5\,_{- 1.7}^{+ 2.1}$    &    --                   &     --                    &    --                     \\ [3ex] 
\\  
J1226+1749a & $-58.5\,_{-0.7}^{+0.6}$ & $11.93\,_{-0.04}^{+0.03}$ & $ 9.2\,_{- 1.0}^{+ 1.3}$    & $  0.4\,_{-1.2}^{+1.1}$ & $11.47\,_{-0.07}^{+0.06}$ & $ 8.9\,_{- 1.9}^{+ 2.5}$  \\ [1ex] 
            & $-19.8\,_{-0.9}^{+1.1}$ & $12.55\,_{-0.03}^{+0.03}$ & $24.3\,_{- 2.4}^{+ 2.9}$    &    --                   &     --                    &    --                     \\ [3ex] 
\\  
J1229+2101a & $-22.7\,_{-7.1}^{+8.7}$ & $12.22\,_{-0.16}^{+0.15}$ & $41.9\,_{-14.7}^{+19.2}$    & $ -1.3\,_{-0.9}^{+0.9}$ & $11.72\,_{-0.08}^{+0.06}$ & $ 8.6\,_{- 1.4}^{+ 1.6}$  \\ [3ex] 
\\  
J1230+2052a & $-31.8\,_{-5.6}^{+3.8}$ & $11.86\,_{-0.12}^{+0.25}$ & $24.5\,_{- 8.1}^{+26.7}$    & $  1.1\,_{-0.3}^{+0.3}$ & $11.77\,_{-0.05}^{+0.07}$ & $ 5.9\,_{- 1.1}^{+ 1.3}$  \\ [1ex] 
            & $ -2.8\,_{-0.9}^{+1.0}$ & $11.70\,_{-0.06}^{+0.07}$ & $ 8.8\,_{- 1.7}^{+ 2.5}$    &    --                   &     --                    &    --                     \\ [3ex] 
\\  
J1231+3719a & $-51.0\,_{-0.7}^{+0.6}$ & $11.78\,_{-0.05}^{+0.04}$ & $ 5.7\,_{- 1.4}^{+ 1.5}$    & $-49.4\,_{-0.9}^{+1.0}$ & $11.29\,_{-0.08}^{+0.09}$ & $ 4.7\,_{- 1.5}^{+ 1.6}$  \\ [1ex] 
            & $-27.0\,_{-0.7}^{+0.6}$ & $12.17\,_{-0.03}^{+0.03}$ & $11.7\,_{- 1.1}^{+ 1.2}$    & $-26.4\,_{-0.3}^{+0.3}$ & $11.77\,_{-0.03}^{+0.03}$ & $ 4.3\,_{- 0.7}^{+ 0.7}$  \\ [1ex] 
            & $  1.6\,_{-3.6}^{+2.3}$ & $11.66\,_{-0.12}^{+0.21}$ & $11.3\,_{- 3.9}^{+ 9.7}$    & $  6.9\,_{-2.7}^{+1.9}$ & $11.32\,_{-0.14}^{+0.14}$ & $ 7.4\,_{- 3.0}^{+ 4.6}$  \\ [3ex] 
\\  
J1234+4437a & $-35.5\,_{-1.8}^{+1.7}$ & $12.54\,_{-0.06}^{+0.12}$ & $26.5\,_{- 4.8}^{+10.1}$    & $-46.0\,_{-0.7}^{+0.7}$ & $11.42\,_{-0.05}^{+0.05}$ & $ 8.2\,_{- 1.6}^{+ 1.7}$  \\ [1ex] 
            & $ -9.9\,_{-0.6}^{+0.6}$ & $12.16\,_{-0.04}^{+0.03}$ & $ 9.2\,_{- 1.0}^{+ 1.3}$    &    --                   &     --                    &    --                     \\ [1ex] 
            & $  8.5\,_{-0.5}^{+0.5}$ & $12.15\,_{-0.03}^{+0.04}$ & $ 8.4\,_{- 1.1}^{+ 1.4}$    &    --                   &     --                    &    --                     \\ [3ex] 
\\  
J1236+2837a & $-36.5\,_{-0.8}^{+0.7}$ & $11.80\,_{-0.05}^{+0.06}$ & $ 8.1\,_{- 1.2}^{+ 1.5}$    & $-18.2\,_{-0.4}^{+0.5}$ & $11.48\,_{-0.04}^{+0.04}$ & $ 3.2\,_{- 1.1}^{+ 1.2}$  \\ [1ex] 
            & $-15.5\,_{-0.6}^{+0.7}$ & $11.86\,_{-0.05}^{+0.05}$ & $ 6.7\,_{- 1.1}^{+ 1.7}$    &    --                   &     --                    &    --                     \\ [1ex] 
            & $ -2.6\,_{-3.9}^{+1.9}$ & $12.00\,_{-0.12}^{+0.19}$ & $16.1\,_{- 5.0}^{+ 9.1}$    &    --                   &     --                    &    --                     \\ [3ex] 
\\  
J1239+2937a & $-40.5\,_{-0.3}^{+0.3}$ & $12.01\,_{-0.02}^{+0.02}$ & $ 8.3\,_{- 0.5}^{+ 0.6}$    & $-39.2\,_{-0.2}^{+0.2}$ & $11.67\,_{-0.03}^{+0.03}$ & $ 2.3\,_{- 0.4}^{+ 0.4}$  \\ [1ex] 
            & $-20.0\,_{-1.5}^{+1.0}$ & $11.71\,_{-0.07}^{+0.12}$ & $10.8\,_{- 2.4}^{+ 5.9}$    &    --                   &     --                    &    --                     \\ [1ex] 
            & $ -0.5\,_{-2.8}^{+1.7}$ & $11.68\,_{-0.12}^{+0.31}$ & $15.5\,_{- 5.3}^{+20.8}$    &    --                   &     --                    &    --                     \\ [3ex] 
\\  
J1240+2746a & $-22.2\,_{-1.1}^{+1.3}$ & $12.26\,_{-0.04}^{+0.04}$ & $24.0\,_{- 3.0}^{+ 4.2}$    & $-28.6\,_{-2.3}^{+2.2}$ & $11.49\,_{-0.06}^{+0.06}$ & $16.9\,_{- 2.7}^{+ 3.6}$  \\ [3ex] 
\\  
J1241+2909a & $-38.8\,_{-1.5}^{+1.0}$ & $12.21\,_{-0.05}^{+0.06}$ & $14.4\,_{- 2.0}^{+ 3.0}$    &    --                   &     --                    &    --                     \\ [1ex] 
            & $ -8.6\,_{-2.4}^{+1.4}$ & $11.75\,_{-0.10}^{+0.14}$ & $10.4\,_{- 2.7}^{+ 4.9}$    &    --                   &     --                    &    --                     \\ [3ex] 
\\  
J1242+2126a & $-30.5\,_{-0.5}^{+0.4}$ & $11.86\,_{-0.03}^{+0.03}$ & $ 5.5\,_{- 0.7}^{+ 0.9}$    & $ -0.8\,_{-0.3}^{+0.3}$ & $12.12\,_{-0.07}^{+0.19}$ & $ 2.3\,_{- 0.6}^{+ 0.6}$  \\ [1ex] 
            & $-11.1\,_{-2.8}^{+2.1}$ & $12.01\,_{-0.09}^{+0.15}$ & $19.2\,_{- 4.6}^{+10.5}$    &    --                   &     --                    &    --                     \\ [3ex] 
\\  
J1243+2842a & $-29.6\,_{-3.7}^{+4.0}$ & $11.76\,_{-0.13}^{+0.17}$ & $17.9\,_{- 5.5}^{+10.6}$    & $ -1.9\,_{-1.4}^{+1.2}$ & $11.59\,_{-0.07}^{+0.07}$ & $10.3\,_{- 1.9}^{+ 4.2}$  \\ [1ex] 
            & $-13.5\,_{-1.1}^{+1.3}$ & $12.02\,_{-0.06}^{+0.06}$ & $17.0\,_{- 2.6}^{+ 4.2}$    &    --                   &     --                    &    --                     \\ [3ex] 
\\  
J1243+2931a & $-39.6\,_{-0.9}^{+1.1}$ & $11.72\,_{-0.05}^{+0.05}$ & $ 9.1\,_{- 1.5}^{+ 2.1}$    & $-21.5\,_{-1.2}^{+2.2}$ & $11.15\,_{-0.08}^{+0.14}$ & $ 7.0\,_{- 2.2}^{+ 4.6}$  \\ [1ex] 
            & $ -9.5\,_{-1.1}^{+1.7}$ & $11.98\,_{-0.08}^{+0.20}$ & $11.5\,_{- 2.6}^{+ 9.2}$    & $ -8.5\,_{-0.2}^{+0.2}$ & $11.55\,_{-0.02}^{+0.02}$ & $ 3.8\,_{- 0.5}^{+ 0.5}$  \\ [3ex] 
\\  
J1244+2926a & $-24.0\,_{-2.4}^{+1.9}$ & $12.17\,_{-0.17}^{+0.22}$ & $12.3\,_{- 4.9}^{+10.1}$    & $-24.2\,_{-0.4}^{+0.4}$ & $11.50\,_{-0.15}^{+0.18}$ & $ 0.7\,_{- 0.1}^{+ 0.5}$  \\ [1ex] 
            & $ -8.9\,_{-0.9}^{+0.7}$ & $12.27\,_{-0.03}^{+0.03}$ & $11.4\,_{- 1.2}^{+ 1.5}$    & $ -6.6\,_{-0.7}^{+0.6}$ & $11.45\,_{-0.04}^{+0.04}$ & $ 7.3\,_{- 1.1}^{+ 1.3}$  \\ [1ex] 
            & $ 19.9\,_{-0.9}^{+1.0}$ & $11.95\,_{-0.06}^{+0.06}$ & $ 9.4\,_{- 1.6}^{+ 2.2}$    &    --                   &     --                    &    --                     \\ [3ex] 
\\  
J1244+2939a & $ -8.4\,_{-0.6}^{+0.5}$ & $12.50\,_{-0.01}^{+0.01}$ & $17.4\,_{- 0.7}^{+ 0.8}$    & $ -6.2\,_{-0.2}^{+0.2}$ & $12.16\,_{-0.02}^{+0.01}$ & $ 5.5\,_{- 0.4}^{+ 0.4}$  \\ [3ex] 
\\  
J1248+0947a & $ -8.9\,_{-0.5}^{+1.3}$ & $11.87\,_{-0.07}^{+0.12}$ & $11.7\,_{- 2.5}^{+ 5.6}$    & $-10.0\,_{-0.3}^{+0.6}$ & $11.61\,_{-0.07}^{+0.06}$ & $ 7.9\,_{- 2.4}^{+ 2.3}$  \\ [3ex] 
\\  
J1251+3849a & $-51.1\,_{-5.5}^{+5.8}$ & $12.28\,_{-0.18}^{+0.17}$ & $35.8\,_{-12.7}^{+19.3}$    &    --                   &     --                    &    --                     \\ [1ex] 
            & $-29.4\,_{-0.8}^{+0.8}$ & $12.19\,_{-0.03}^{+0.03}$ & $14.8\,_{- 1.3}^{+ 1.7}$    &    --                   &     --                    &    --                     \\ [1ex] 
            & $ -6.2\,_{-0.5}^{+0.6}$ & $11.93\,_{-0.04}^{+0.04}$ & $ 7.5\,_{- 0.9}^{+ 1.1}$    &    --                   &     --                    &    --                     \\ [3ex] 
\\  
J1255+0924a & $-19.4\,_{-1.1}^{+2.0}$ & $11.92\,_{-0.10}^{+0.15}$ & $13.6\,_{- 3.2}^{+ 7.8}$    & $-15.4\,_{-3.3}^{+4.7}$ & $12.02\,_{-0.21}^{+0.23}$ & $31.4\,_{-13.4}^{+23.6}$  \\ [1ex] 
            & $ -6.2\,_{-0.5}^{+0.5}$ & $11.75\,_{-0.04}^{+0.04}$ & $ 6.4\,_{- 1.2}^{+ 1.4}$    & $ -6.4\,_{-0.4}^{+0.4}$ & $11.41\,_{-0.04}^{+0.04}$ & $ 2.8\,_{- 1.1}^{+ 1.3}$  \\ [3ex] 
\\  
J1258+1939a & $-13.7\,_{-2.6}^{+4.3}$ & $12.22\,_{-0.07}^{+0.10}$ & $26.0\,_{- 4.2}^{+ 8.0}$    & $  0.0\,_{-0.5}^{+0.6}$ & $11.44\,_{-0.07}^{+0.07}$ & $ 2.7\,_{- 1.2}^{+ 1.5}$  \\ [1ex] 
            & $  0.9\,_{-1.1}^{+0.9}$ & $11.92\,_{-0.05}^{+0.07}$ & $ 9.8\,_{- 1.9}^{+ 2.9}$    &    --                   &     --                    &    --                     \\ [1ex] 
            & $ 40.8\,_{-0.9}^{+0.9}$ & $11.55\,_{-0.07}^{+0.06}$ & $ 5.5\,_{- 1.5}^{+ 1.6}$    &    --                   &     --                    &    --                     \\ [3ex] 
\\  
J1259+1552a & $-19.0\,_{-0.9}^{+1.0}$ & $12.24\,_{-0.08}^{+0.10}$ & $ 6.7\,_{- 2.4}^{+ 3.1}$    &    --                   &     --                    &    --                     \\ [1ex] 
            & $ -1.4\,_{-5.9}^{+8.0}$ & $12.37\,_{-0.23}^{+0.24}$ & $30.6\,_{-14.3}^{+28.8}$    &    --                   &     --                    &    --                     \\ [3ex] 
\\  
J1306+2003a & $-13.7\,_{-1.6}^{+3.1}$ & $12.17\,_{-0.08}^{+0.19}$ & $13.5\,_{- 3.7}^{+ 9.1}$    & $  7.8\,_{-0.2}^{+0.2}$ & $11.97\,_{-0.15}^{+0.22}$ & $ 1.4\,_{- 0.3}^{+ 1.1}$  \\ [1ex] 
            & $  1.9\,_{-0.8}^{+0.6}$ & $12.34\,_{-0.04}^{+0.05}$ & $10.5\,_{- 1.5}^{+ 1.9}$    &    --                   &     --                    &    --                     \\ [3ex] 
\\  
J1307+2224a &    --                   &     --                    &   --                        & $  0.8\,_{-3.6}^{+1.9}$ & $11.67\,_{-0.23}^{+0.44}$ & $13.8\,_{- 7.6}^{+27.0}$  \\ [3ex]   
\\  
J1310+2514a & $-33.8\,_{-0.9}^{+0.9}$ & $11.82\,_{-0.07}^{+0.07}$ & $ 6.3\,_{- 1.4}^{+ 1.8}$    &    --                   &     --                    &    --                     \\ [1ex] 
            & $ -9.8\,_{-5.7}^{+3.3}$ & $12.15\,_{-0.12}^{+0.23}$ & $20.9\,_{- 6.6}^{+18.7}$    &    --                   &     --                    &    --                     \\ [3ex] 
\\  
J1314+1751a &    --                   &     --                    &   --                        & $ -2.2\,_{-0.3}^{+0.3}$ & $11.77\,_{-0.02}^{+0.03}$ & $ 3.7\,_{- 0.7}^{+ 0.7}$  \\ [3ex]   
\\  
J1324+2038a & $-17.9\,_{-1.0}^{+1.4}$ & $12.20\,_{-0.05}^{+0.07}$ & $14.1\,_{- 2.0}^{+ 3.6}$    & $ 13.8\,_{-0.9}^{+1.1}$ & $11.32\,_{-0.08}^{+0.37}$ & $ 4.2\,_{- 2.0}^{+13.3}$  \\ [1ex] 
            & $ 15.8\,_{-0.3}^{+0.3}$ & $12.08\,_{-0.04}^{+0.07}$ & $ 3.1\,_{- 0.8}^{+ 0.7}$    &    --                   &     --                    &    --                     \\ [3ex] 
\\  
J1324+2418a & $-24.5\,_{-0.8}^{+0.7}$ & $11.97\,_{-0.05}^{+0.04}$ & $ 6.6\,_{- 1.3}^{+ 1.4}$    & $-23.2\,_{-0.3}^{+0.3}$ & $11.72\,_{-0.03}^{+0.02}$ & $ 4.5\,_{- 0.5}^{+ 0.5}$  \\ [3ex] 
\\  
J1325+2232aa & $-14.8\,_{-1.6}^{+2.1}$ & $12.36\,_{-0.07}^{+0.06}$ & $20.7\,_{- 3.1}^{+ 3.5}$   & $-26.1\,_{-0.4}^{+0.4}$ & $11.71\,_{-0.04}^{+0.03}$ & $ 6.3\,_{- 0.9}^{+ 0.8}$  \\ [3ex] 
\\  
J1325+2232ab & $-17.3\,_{-0.3}^{+0.3}$ & $12.19\,_{-0.02}^{+0.02}$ & $ 9.7\,_{- 0.5}^{+ 0.6}$   & $-18.2\,_{-1.6}^{+1.4}$ & $11.29\,_{-0.11}^{+0.16}$ & $ 8.0\,_{- 3.1}^{+ 8.0}$  \\ [3ex] 
\\  
J1332+2054a & $-11.5\,_{-1.1}^{+1.1}$ & $11.57\,_{-0.03}^{+0.03}$ & $ 3.2\,_{- 0.8}^{+ 0.8}$    & $-26.7\,_{-0.5}^{+0.5}$ & $11.29\,_{-0.10}^{+0.17}$ & $ 0.9\,_{- 0.4}^{+ 1.5}$  \\ [1ex]   
            &    --                   &     --                    &   --                        & $- 0.6\,_{-0.3}^{+0.3}$ & $11.57\,_{-0.03}^{+0.03}$ & $ 3.2\,_{- 0.8}^{+ 0.8}$  \\ [3ex] 
\\  
J1335+2820a & $-24.2\,_{-0.6}^{+0.5}$ & $12.35\,_{-0.03}^{+0.03}$ & $11.3\,_{- 0.9}^{+ 0.9}$    &    --                   &     --                    &    --                     \\ [3ex] 
\\  
J1338+2345a &    --                   &     --                    &   --            &    --                   &     --                    &    --                     \\ [3ex]   
\\  
J1341+2801a & $-33.7\,_{-0.9}^{+1.0}$ & $12.51\,_{-0.02}^{+0.02}$ & $28.1\,_{- 1.6}^{+ 2.4}$    &    --                   &     --                    &    --                     \\ [3ex] 
\\  
J1341+2806a & $-72.6\,_{-3.5}^{+4.4}$ & $12.42\,_{-0.13}^{+0.15}$ & $35.4\,_{-10.0}^{+19.1}$    &    --                   &     --                    &    --                     \\ [1ex] 
            & $-26.2\,_{-2.0}^{+1.9}$ & $12.46\,_{-0.03}^{+0.04}$ & $35.8\,_{- 3.8}^{+ 5.2}$    &    --                   &     --                    &    --                     \\ [3ex]   
\\  
J1341+2823a & $-31.1\,_{-0.3}^{+0.3}$ & $12.17\,_{-0.02}^{+0.02}$ & $ 9.0\,_{- 0.5}^{+ 0.5}$    & $-32.2\,_{-0.2}^{+0.1}$ & $12.00\,_{-0.11}^{+0.21}$ & $ 0.9\,_{- 0.2}^{+ 0.2}$    \\ [3ex]  
\\  
J1341+2824a & $-28.0\,_{-0.5}^{+0.6}$ & $12.32\,_{-0.02}^{+0.02}$ & $14.9\,_{- 0.8}^{+ 0.9}$    &    --                   &     --                    &    --                     \\ [3ex] 
\\  
J1341+2829a & $-27.5\,_{-0.4}^{+0.4}$ & $12.39\,_{-0.03}^{+0.02}$ & $ 7.6\,_{- 0.6}^{+ 0.6}$    &    --                   &     --                    &    --                     \\ [3ex] 
\\  
J1342+2828a & $-25.2\,_{-0.5}^{+0.6}$ & $12.22\,_{-0.02}^{+0.02}$ & $12.8\,_{- 0.9}^{+ 1.0}$    &    --                   &     --                    &    --                     \\ [3ex] 
\\  
J1344+1842a & $-24.2\,_{-0.2}^{+0.2}$ & $12.38\,_{-0.01}^{+0.01}$ & $14.1\,_{- 0.5}^{+ 0.5}$    & $-24.8\,_{-1.2}^{+2.0}$ & $11.24\,_{-0.09}^{+0.14}$ & $ 8.6\,_{- 3.1}^{+ 7.7}$  \\ [1ex] 
            & $ -5.2\,_{-0.6}^{+0.5}$ & $12.03\,_{-0.02}^{+0.02}$ & $11.2\,_{- 0.8}^{+ 0.9}$    & $ -3.5\,_{-0.2}^{+0.2}$ & $11.68\,_{-0.02}^{+0.02}$ & $ 2.7\,_{- 0.5}^{+ 0.6}$  \\ [3ex] 
\\  
J1347+1811a & $ -9.4\,_{-0.6}^{+0.7}$ & $12.49\,_{-0.03}^{+0.03}$ & $20.4\,_{- 2.0}^{+ 2.0}$    & $ -1.8\,_{-0.3}^{+0.3}$ & $11.74\,_{-0.02}^{+0.02}$ & $ 6.6\,_{- 0.6}^{+ 0.7}$  \\ [3ex] 
\\  
J1413+5621a & $-45.4\,_{-0.2}^{+0.2}$ & $12.04\,_{-0.02}^{+0.02}$ & $ 4.4\,_{- 1.0}^{+ 0.9}$    & $-45.6\,_{-0.1}^{+0.1}$ & $12.07\,_{-0.06}^{+0.07}$ & $ 1.5\,_{- 0.1}^{+ 0.2}$  \\ [1ex] 
            & $ -4.8\,_{-1.1}^{+1.0}$ & $11.88\,_{-0.04}^{+0.04}$ & $12.6\,_{- 1.6}^{+ 1.9}$    & $ -4.5\,_{-1.2}^{+1.3}$ & $11.25\,_{-0.07}^{+0.07}$ & $10.3\,_{- 2.2}^{+ 2.7}$  \\ [3ex] 
\\  
J1415+3716a & $-36.9\,_{-0.8}^{+0.8}$ & $11.57\,_{-0.04}^{+0.05}$ & $ 6.9\,_{- 1.2}^{+ 1.5}$    & $ -5.9\,_{-1.7}^{+1.4}$ & $11.11\,_{-0.10}^{+0.10}$ & $ 6.0\,_{- 2.6}^{+ 3.2}$  \\ [1ex] 
            & $ -8.7\,_{-1.0}^{+1.1}$ & $11.98\,_{-0.03}^{+0.03}$ & $16.6\,_{- 1.9}^{+ 1.9}$    &    --                   &     --                    &    --                     \\ [3ex] 
\\  
J1420+5520a & $-11.6\,_{-0.5}^{+0.6}$ & $11.82\,_{-0.03}^{+0.04}$ & $ 7.9\,_{- 1.1}^{+ 1.6}$    &    --                   &     --                    &    --                     \\ [3ex] 
\\  
J1527+4027a & $-72.4\,_{-0.3}^{+0.3}$ & $11.57\,_{-0.02}^{+0.02}$ & $ 5.4\,_{- 0.6}^{+ 0.6}$    & $ -3.0\,_{-0.1}^{+0.1}$ & $11.55\,_{-0.02}^{+0.02}$ & $ 2.6\,_{- 0.5}^{+ 0.4}$    \\ [3ex]   
            & $-54.7\,_{-0.7}^{+0.5}$ & $11.31\,_{-0.05}^{+0.06}$ & $ 5.0\,_{- 1.2}^{+ 1.6}$    &    --                   &     --                    &    --                     \\ [1ex] 
            & $-28.2\,_{-0.3}^{+0.3}$ & $11.46\,_{-0.03}^{+0.03}$ & $ 4.5\,_{- 0.7}^{+ 0.7}$    &    --                   &     --                    &    --                     \\ [1ex]     
            & $ -6.5\,_{-0.2}^{+0.2}$ & $12.08\,_{-0.01}^{+0.01}$ & $ 9.6\,_{- 0.3}^{+ 0.3}$    &    --                   &     --                    &    --                     \\ [3ex] 
\\
J1534+5015a & $-51.3\,_{-0.1}^{+0.1}$ & $12.02\,_{-0.01}^{+0.01}$ & $ 5.1\,_{- 0.2}^{+ 0.2}$    & $-30.9\,_{-0.2}^{+0.2}$ & $11.81\,_{-0.02}^{+0.01}$ & $4.4\,_{- 0.3}^{+ 0.3}$    \\ [3ex]  
          & $-28.4\,_{-0.04}^{+0.04}$ & $12.90\,_{-0.03}^{+0.03}$ & $ 4.1\,_{- 0.1}^{+ 0.1}$    & $ -0.6\,_{-0.2}^{+0.2}$ & $11.66\,_{-0.02}^{+0.03}$ & $2.6\,_{- 0.6}^{+ 0.6}$    \\ [1ex] 
            & $ -4.8\,_{-0.3}^{+0.3}$ & $12.26\,_{-0.01}^{+0.01}$ & $15.9\,_{- 0.4}^{+ 0.5}$    &    --                   &     --                    &    --                     \\ [3ex] 
\enddata
\tablecomments{Properties of all CaII and NaI gas absorption features measured along BHB sightlines. $v_{\rm LSR}$ is the measured radial velocity corrected for the Local Standard of Rest in the direction of that sightline. logN is column density, and $b_{\rm D}$ is the Doppler b parameter of the absorption feature. Errors are the upper and lower quartiles determined by the MCMC line-fitting algorithm. }
\label{table:mcmclist}
\end{deluxetable*}


\clearpage

\appendix

\counterwithin{figure}{section}
\section{Milky Way ISM Component Results}
\label{appendix:MW}

\subsection{Column Density}
\begin{figure*}[h]
    \centering
    \includegraphics[width=\textwidth]{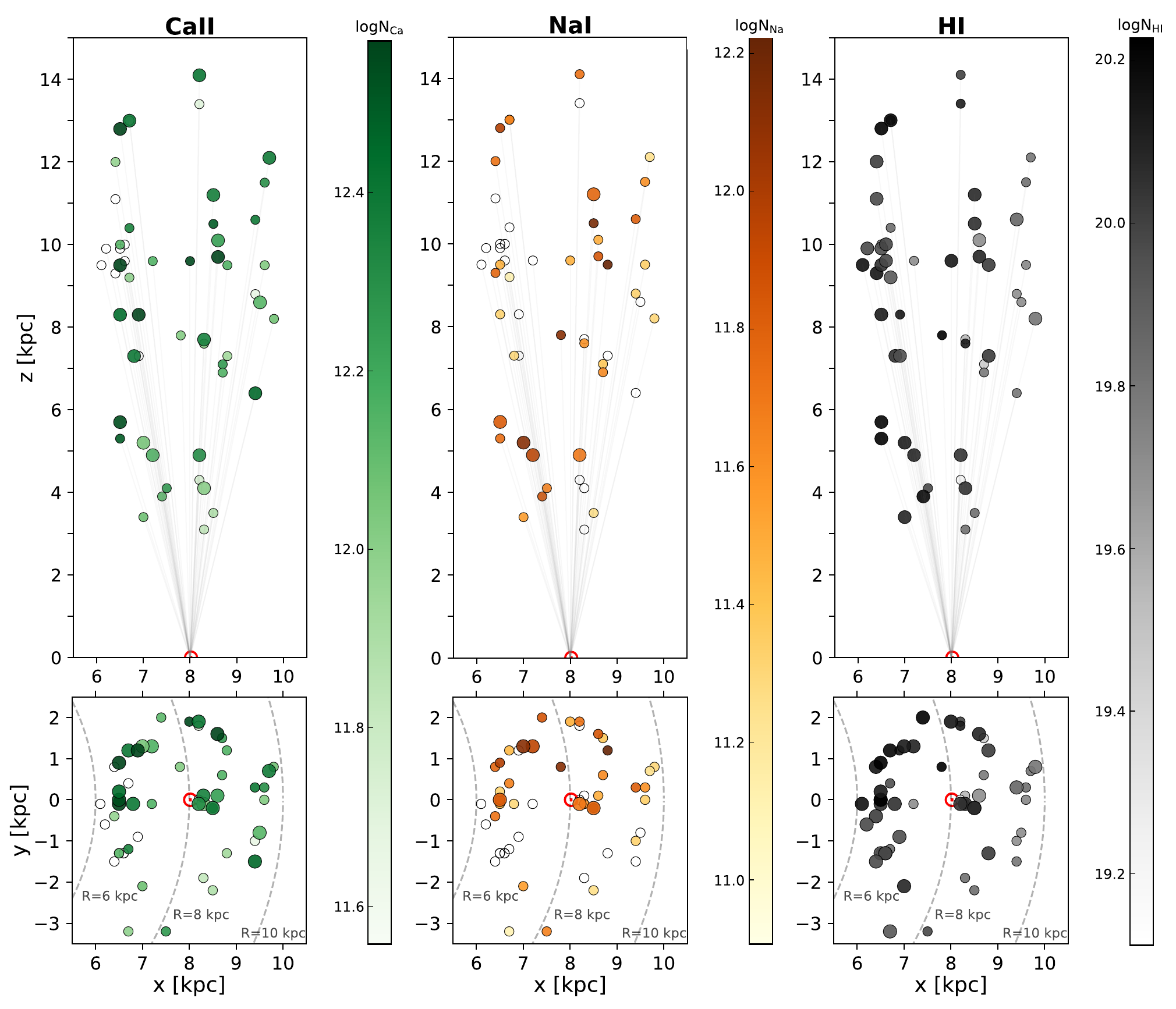}
    \caption{{\sc ISM component column density measurements.} Column densities for the ISM component (Milky Way ISM gas, $|v|<25$ km/s). Markers correspond to background BHB sightlines projected onto the $x$-$y$ plane of the Milky Way disk. The coordinates for each BHB are plotted in physical kpc units in the plane of the Milky Way; thus any inferred distance to gas absorption along a sightline is an upper limit. Color indicates column density measurements for CaII and NaI in absorption and HI in 21-cm emission. Larger markers indicate sightlines along which more than one component was detected, and their colors represent the mean column density of those components. Empty circles denote nondetections. The Sun is marked by a red solar symbol, and black dashes denote lines of constant radius from Galactic center.}
    \label{fig:MWcol}
\end{figure*}

ISM component column densities are shown in Figure \ref{fig:MWcol}, which can be compared directly with the IV component in Figure \ref{fig:IVcol}. Higher CaII and NaI column density detections are grouped together at positive $y$ values (the direction of Galactic rotation), in constrast to the IV component which has high CaII and NaI column density detections clustered together at positive $x$ values (the direction away from Galactic center).
The pattern for HI column densities, which are higher at $x < 8$ kpc, do not match CaII and NaI for ISM gas as they do for IV gas. The Sun, at $x = 8$ kpc, is nestled just outside the the Sagittarius spiral arm of the Milky Way, and the elevated HI column density measurements we measure in that direction can be explained by this feature \citep{vallee2016,vallee2017}.

\subsection{Kinematics}
\begin{figure*}
    \centering
    \includegraphics[width=\textwidth]{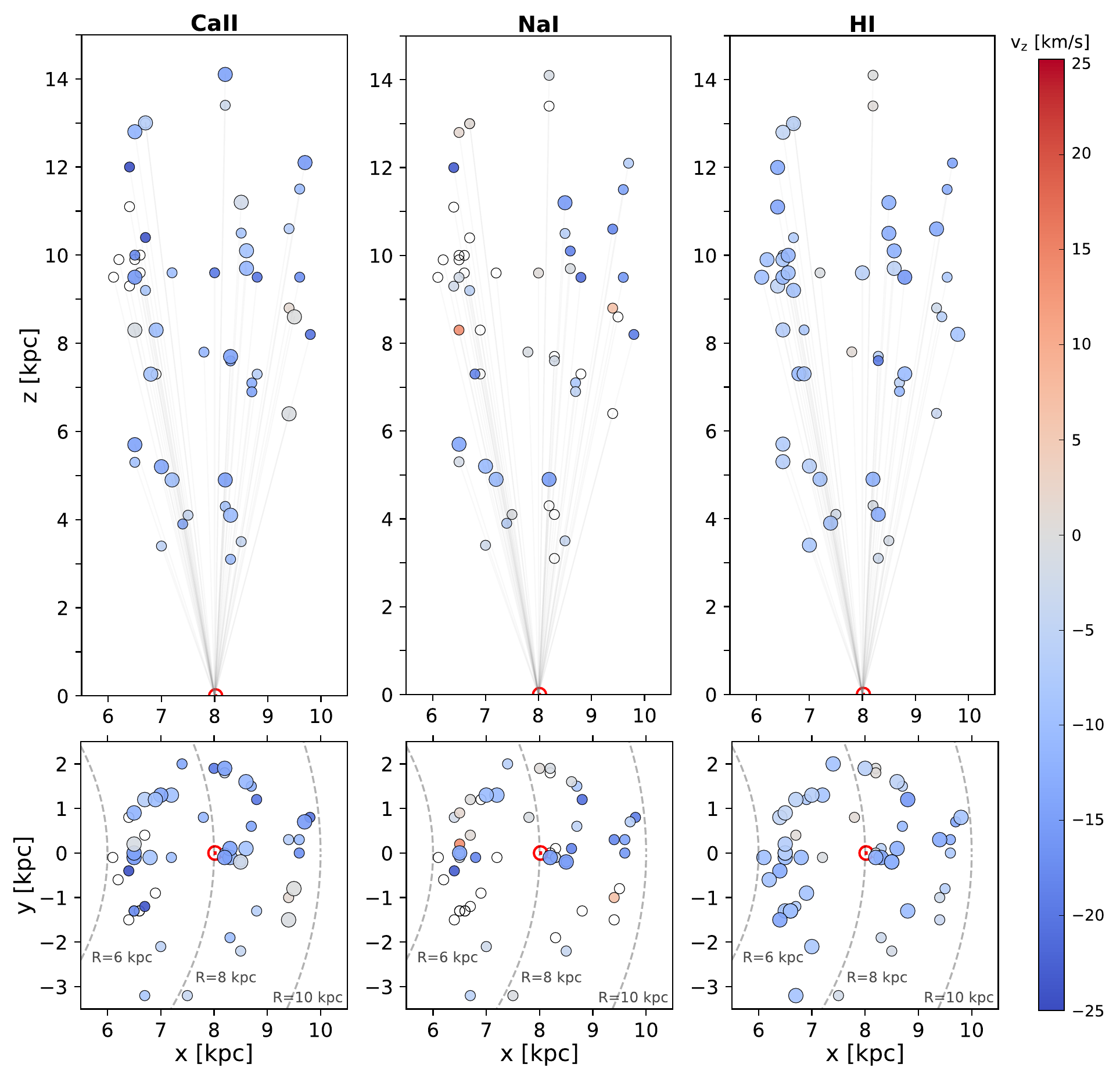}
    \caption{{\sc ISM component velocity measurements.} Velocities for the ISM component (Milky Way ISM gas, $|v|<25$ km/s). Markers correspond to background BHB sightlines projected onto the $x$-$y$ plane of the Milky Way disk. The coordinates for each BHB are plotted in physical kpc units in the plane of the Milky Way; thus any inferred distance to gas absorption along a sightline is an upper limit. Color indicates the component of radial velocity measurements perpendicular to the disk ($v_{\rm LSR}$ sin$b$) for CaII and NaI in absorption and HI in 21-cm emission. Larger markers indicate sightlines along which more than one component was detected, and their colors represent the mean velocity of those components weighted by column density. Empty circles denote nondetections. The Sun is marked by a red solar symbol, and black dashes denote lines of constant radius from Galactic center. In contrast to the `IV' component, the CaII and NaI ISM component does not exhibit any spatial velocity gradient across space.}
    \label{fig:MWvel}
\end{figure*}

Figure \ref{fig:MWvel} shows 
ISM gas component, which again is directly comparable to the IV component results shown in Figure \ref{fig:IVvel}. Detections of Milky Way disk gas (Figure \ref{fig:MWvel}) are relatively uniform and show no clear spatial trends in velocity, although most of the gas is moving at negative velocities in the same direction as accreting gas in the IV component. This is somewhat unexpected for normal gas in the disk, which should have an average $z$-component velocity close to zero relative to the local standard of rest. Therefore, it is possible that these metals are not associated with normal ISM gas and are either slow-moving extraplanar gas that was not separated from disk gas by our $|v|$ $<$ 25 km/s velocity cut, or gas affected by some other phenomenon within the disk \citep{zheng2015}.

\clearpage

\clearpage


\counterwithin{figure}{section}
\section{Absorption Features \& Line Profile Fits}
\label{appendix:linefits}

\begin{figure*}[h]
    \centering
    \includegraphics[width=0.9\textwidth]{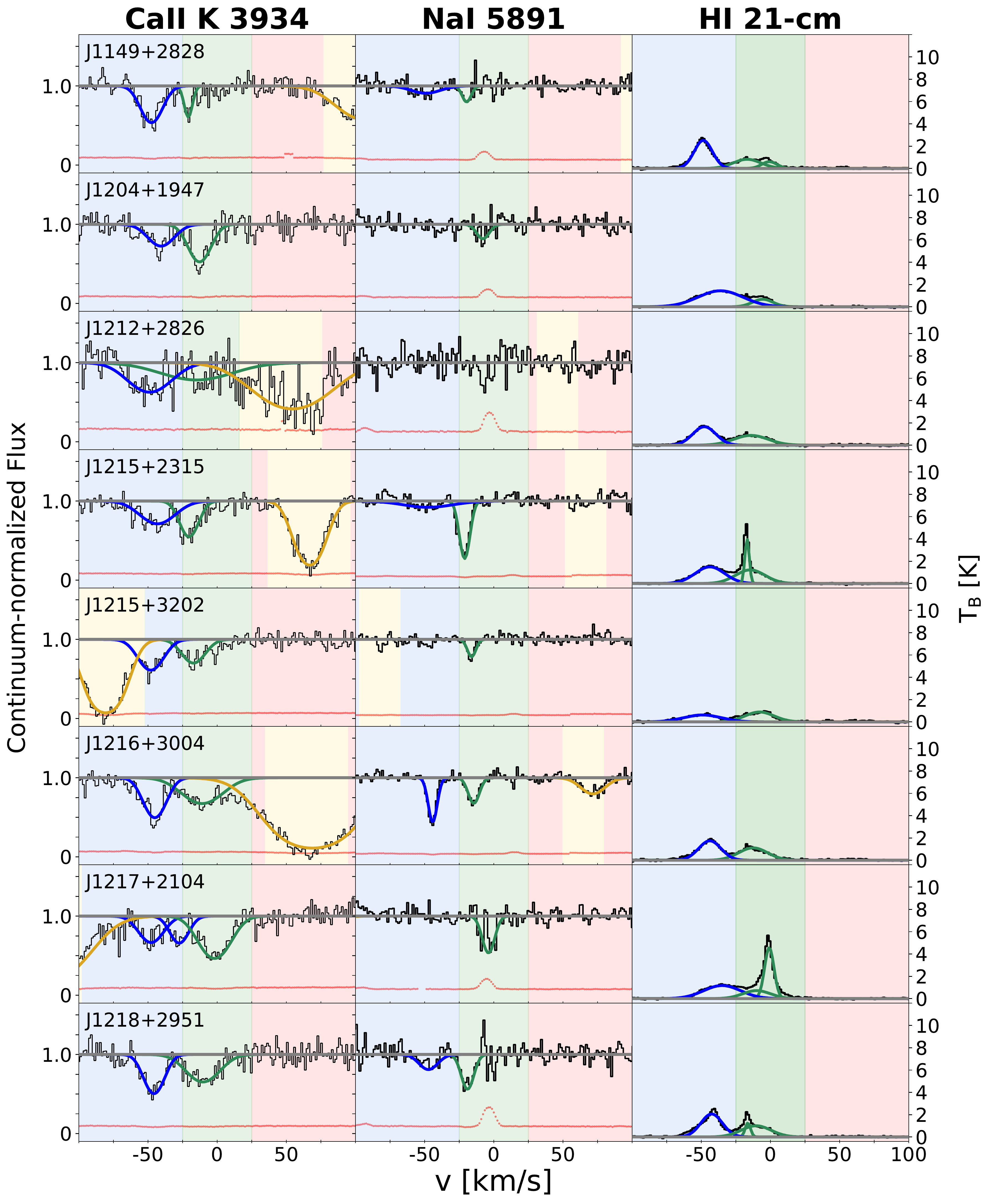}
    \caption{ }
\end{figure*}

\begin{figure*}
    \centering
    \includegraphics[width=0.9\textwidth]{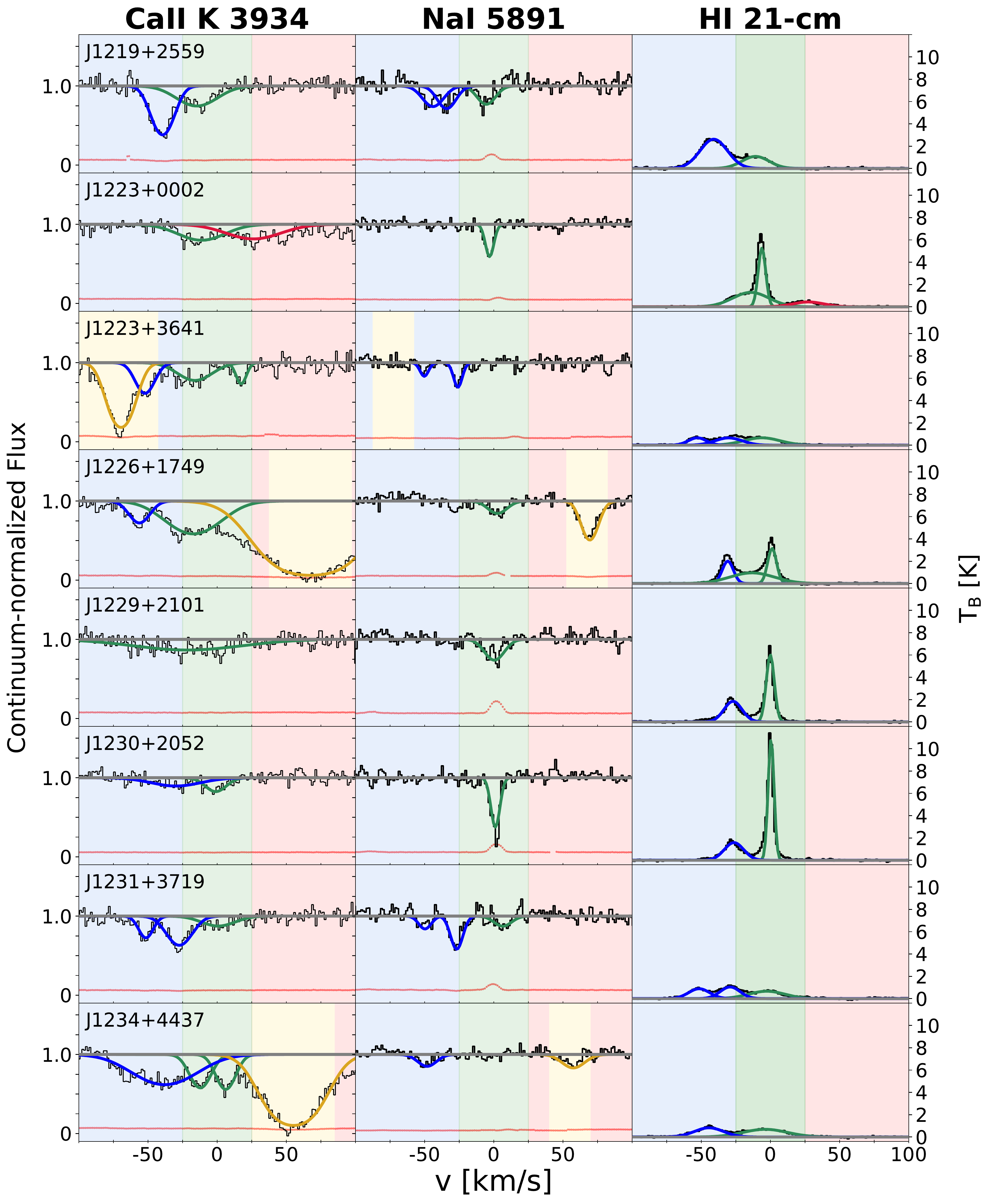}
    \caption{ }
\end{figure*}

\begin{figure*}
    \centering
    \includegraphics[width=0.9\textwidth]{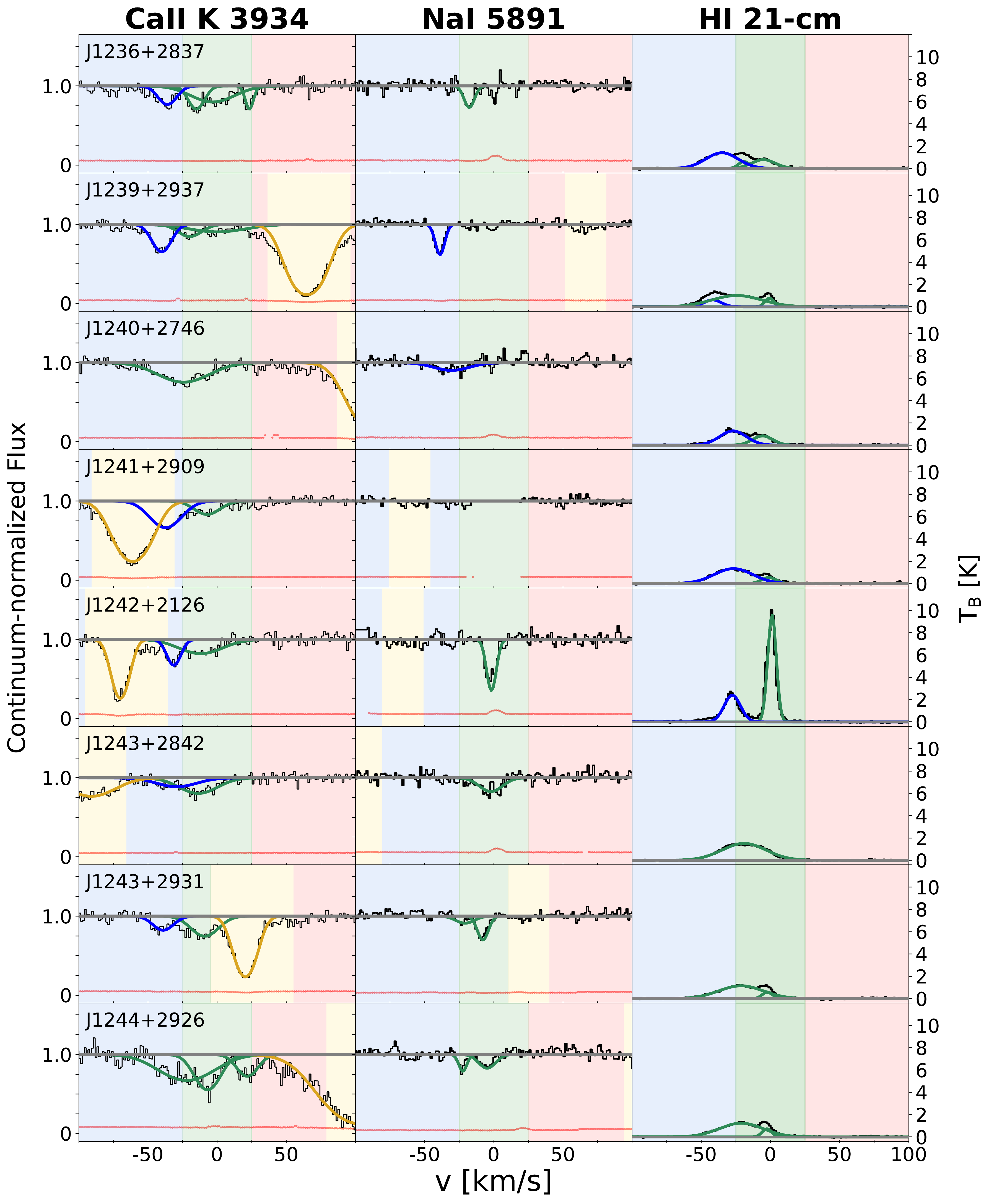}
    \caption{ }
\end{figure*}

\begin{figure*}
    \centering
    \includegraphics[width=0.9\textwidth]{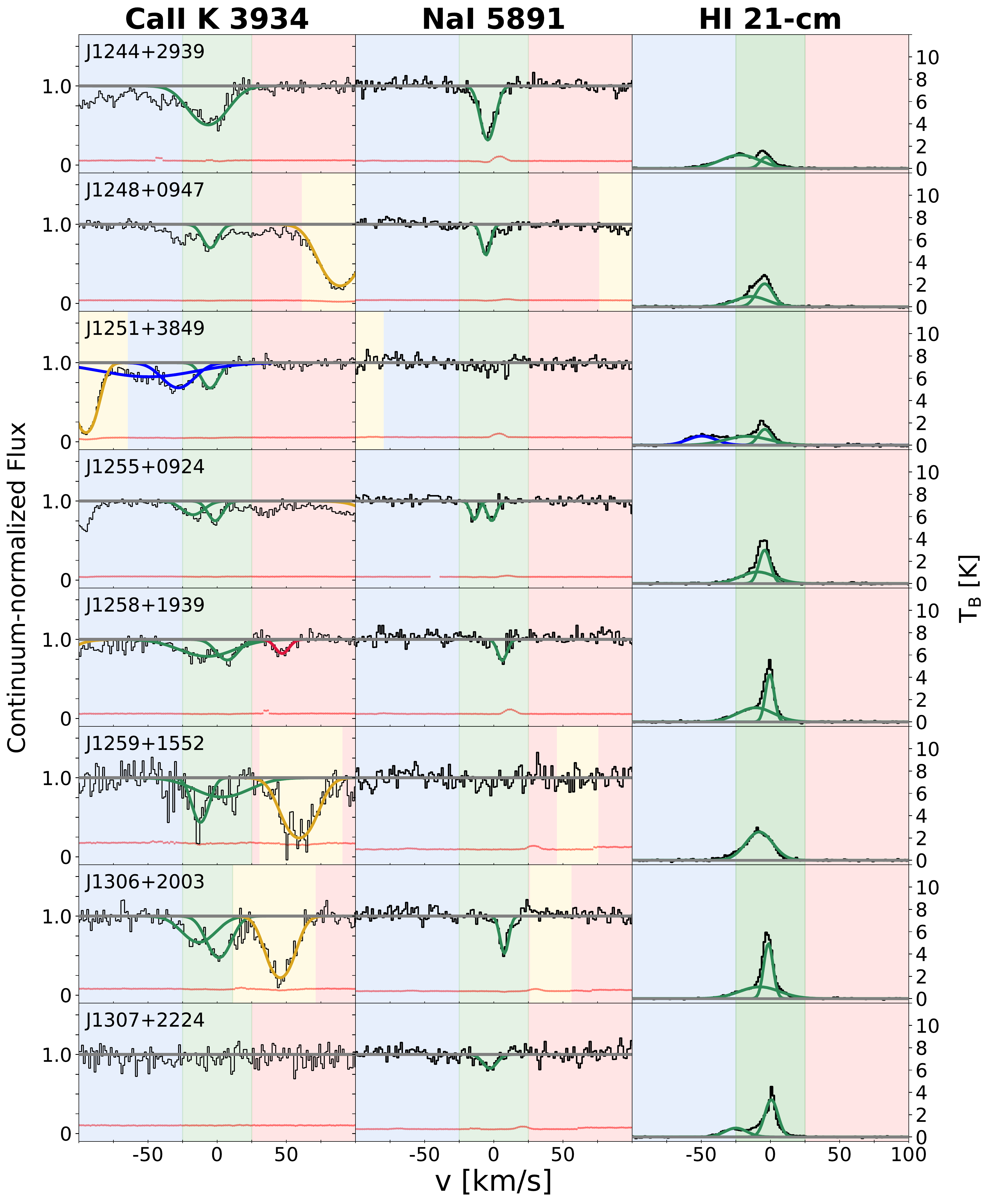}
    \caption{ }
\end{figure*}

\begin{figure*}
    \centering
    \includegraphics[width=0.9\textwidth]{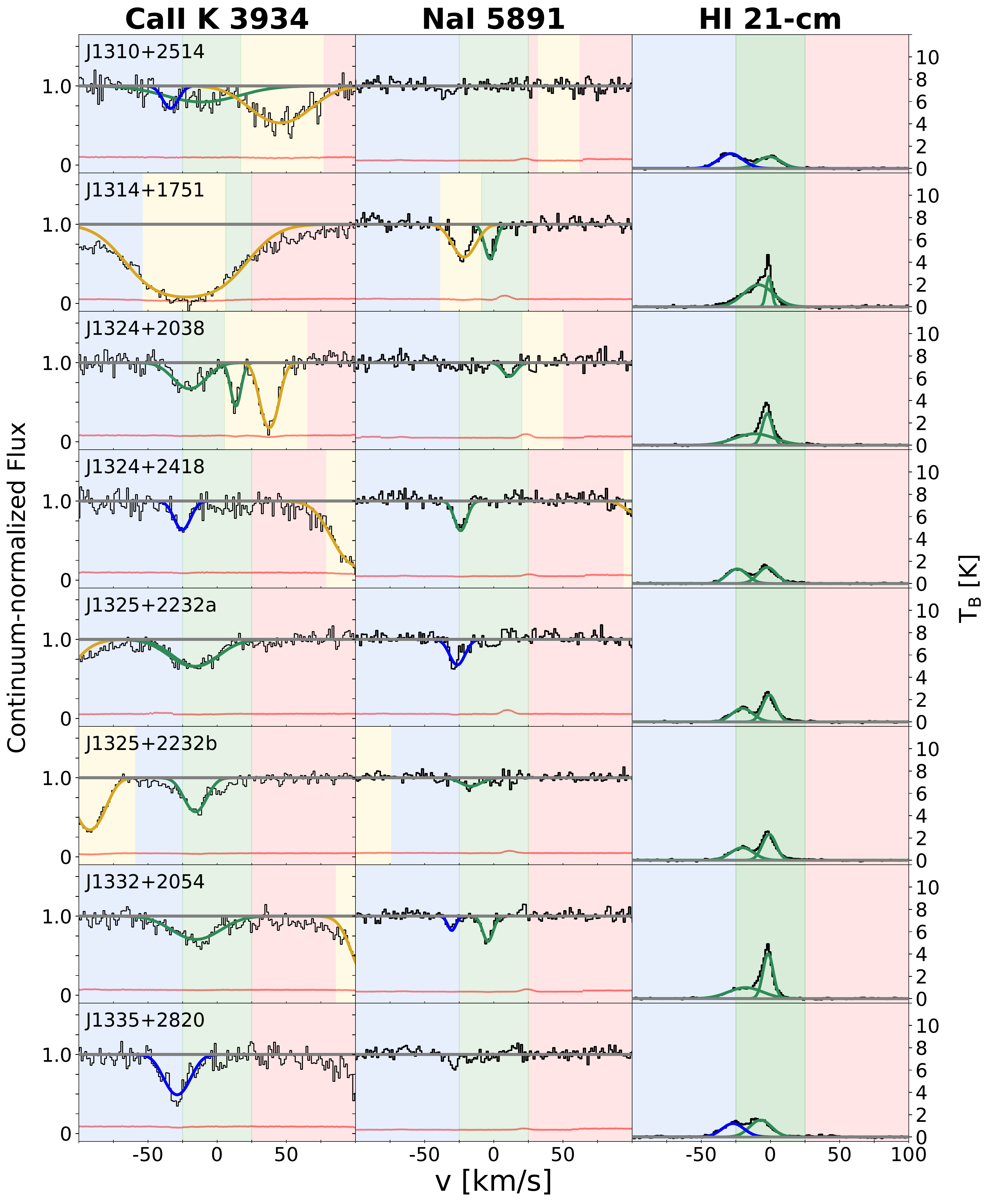}
    \caption{ }
\end{figure*}

\begin{figure*}
    \centering
    \includegraphics[width=0.9\textwidth]{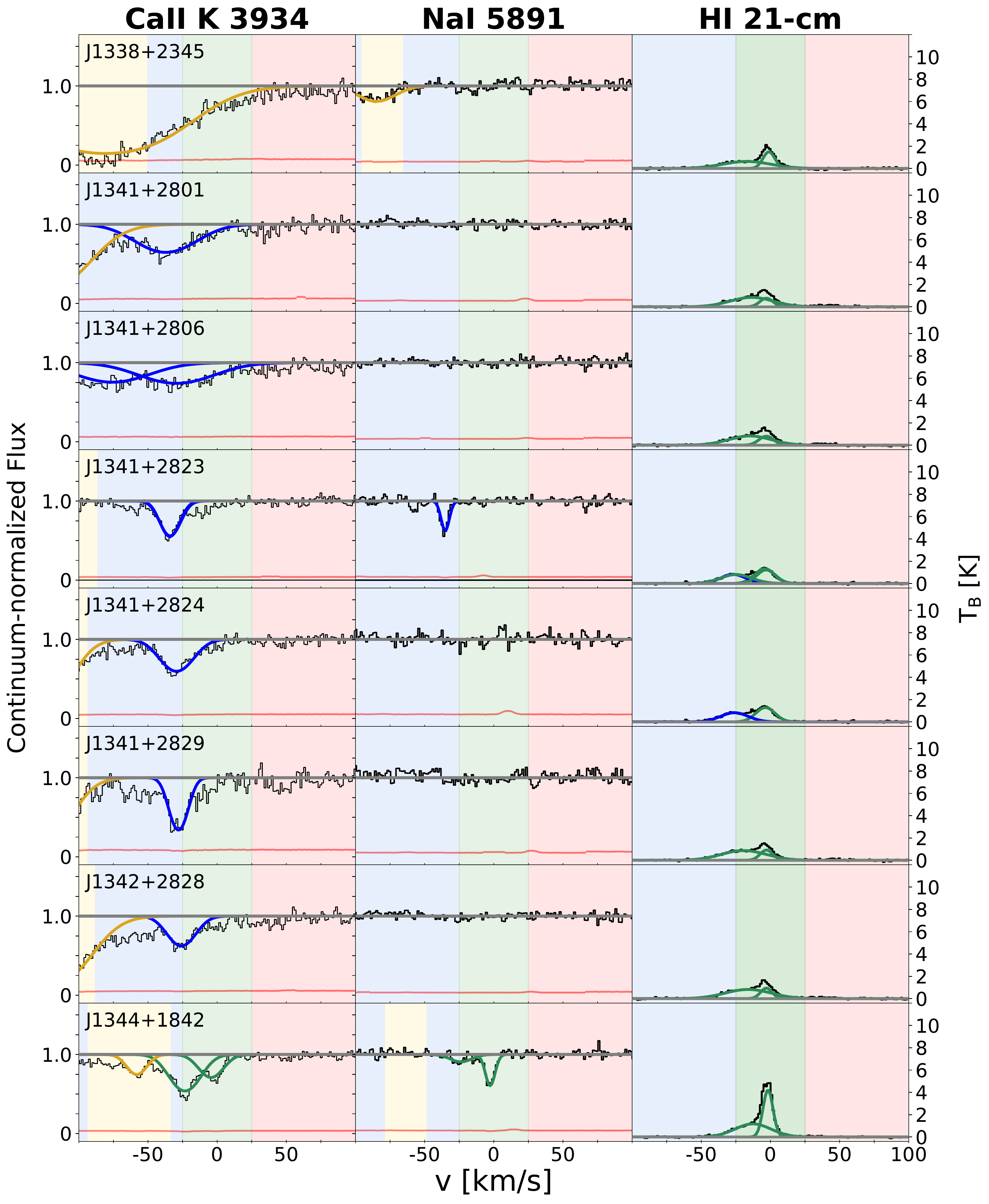}
    \caption{ }
\end{figure*}

\begin{figure*}
    \centering
    \includegraphics[width=0.9\textwidth]{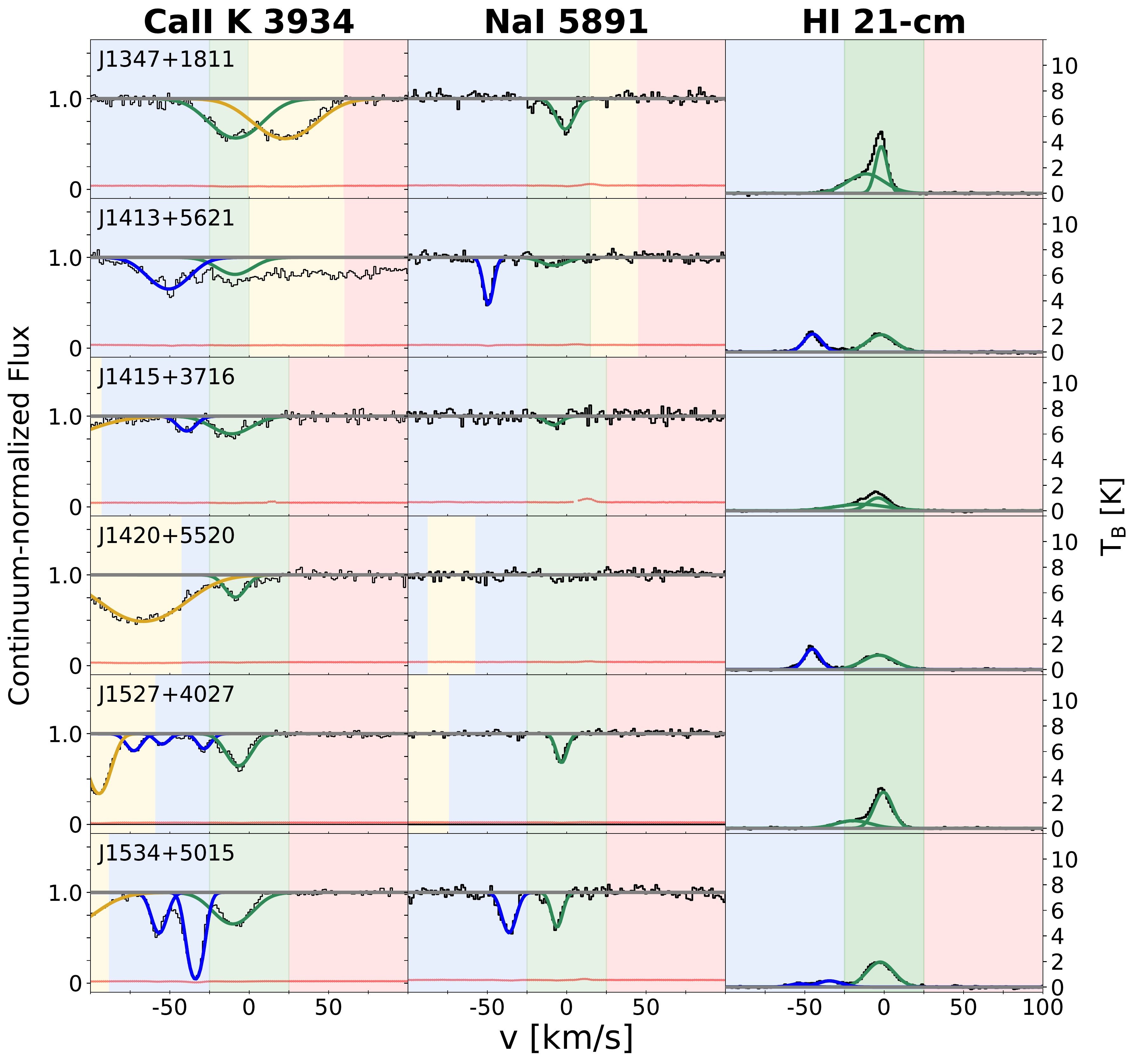}
    \caption{ }
\end{figure*}

\end{document}